# Energy vs. density on paths toward more exact density functionals

Kasper P. Kepp*

*Technical University of Denmark, DTU Chemistry, Building 206, 2800 Kgs. Lyngby, DK – Denmark.* *Phone: +045 45 25 24 09. E-mail: kpj@kemi.dtu.dk

## Abstract

Recently, the progression toward more exact density functional theory has been questioned, implying a need for more formal ways to systematically measure progress, i.e. a path. Here I use the Hohenberg-Kohn theorems and the definition of normality by Burke et al. to define a path toward exactness and straying from the path by separating errors in $\rho$ and $E[\rho]$. A consistent path toward exactness involves minimizing both errors. Second, a suitably diverse test set of trial densities $\rho'$ can be used to estimate the significance of errors in $\rho$ without knowing the exact densities which are often computationally inaccessible. To illustrate this, the systems previously studied by Medvedev et al., the first ionization energies of atoms with $Z = 1$ to 10, the ionization energy of water, and the bond dissociation energies of five diatomic molecules were investigated and benchmarked against CCSD(T)/aug-cc-pV5Z. A test set of four functionals of distinct designs was used: B3LYP, PBE, M06, and S-VWN. For atomic cations *regardless* of charge and compactness up to $Z = 10$, the energy effects of variations in $\rho$ are < 4 kJ/mol (chemical accuracy) defined here as normal, even though these four functionals ranked very differently in the previous test. Thus, the off-path behavior for such cations is energy-wise insignificant and in fact, indeterminate because of noise from other errors. An interesting oscillating behavior in the density sensitivity is observed vs. $Z$, explained by orbital occupation effects. Finally, it is shown that even large normal problems such as the Co-C bond energy of cobalamins can use simpler (e.g. PBE) trial densities to drastically speed up computation by loss of a few kJ/mol in accuracy.

**Introduction**

A major challenge in current theoretical chemistry is the development of new, more generally applicable and accurate functionals, because all functionals lack *general* high accuracy: Some are accurate for some electronic systems, others are for other systems, and the search for "universal" accuracy is thus ongoing[1–6]. An exact functional has a negligible error (i.e. smaller than the uncertainty in standard approximations beyond the functional itself) in *both* the density ρ and its derived properties, notably the associated energy $E[ρ]$, for a given electronic system (*numerically exact*); in addition it should fulfil all fundamental exact physical conditions (*fundamentally exact*). The *universal* functional is the functional that is exact for *any* system and for any physical observable[7]. $E[ρ]$ itself is not observable, except in principle by reverse mapping from an experimental ρ obtained by e.g. X-ray diffraction using the universal functional, via the one-to-one correspondence[4]. Thus, accuracy of a functional is typically assessed by the errors in derived energies of the type $\Delta E = E_2[ρ_2] – E_1[ρ_1]$ where $\Delta E$ represents a chemical conversion whose energy change is observable, e.g. the ionization potential, with a target accuracy of the order of ~4 kJ/mol (chemical accuracy).

In a recent report[2] it was claimed that modern density functionals are becoming less exact. Specifically, the paper indicated that some relatively new, highly parameterized functionals produce less accurate ρ for charged $1s^2$ and $1s^22s^2$ atomic ions, although they generally perform well for derived energies $\Delta E[ρ]$. From this observation it was concluded that modern DFT is straying from the path toward exactness[2]. The debate goes to the very center of the theory, specifically the Hohenberg-Kohn correspondence between the electron density ρ and its derived properties[4]. Functionals can be accurate for one property but inaccurate for another property for the same system[5,8,9]. To speak of a path toward exactness, errors must be assessed not only in ρ but also $E[ρ]$ *for the same systems*, because exactness applies to a specific system[10]. In fact, the errors of ρ and Δ$E$[ρ] are commonly not related even for the same systems except in very distinct simple cases such as, incidentally, the 2s$^2$ double ionization energies of these particular systems[10], and many functionals produce decent ρ (in terms of their ability to serve as trial densities for more exact functionals) but inaccurate Δ$E$[ρ][11]. These two properties are particularly central to DFT because they feature in the Hohenberg-Kohn theorems as having a one-to-one correspondence[4], where ρ obeys the variational principle and is N- and v-representable[3,12].

One question is whether there is a monotonic path toward exactness as ρ → ρ$_{exact}$ and $E$[ρ] → $E_{exact}$[ρ $_{exact}$]. To visualize this one can plot errors in ρ and $E$[ρ] in a diagram such as **Figure 1A**, where a path is given when and only when ρ → ρ$_{exact}$ *and* $E$[ρ] → $E_{exact}$[ρ $_{exact}$] *for the same electronic system*. In the case of no improvement in ρ but only in $E$[ρ], we move horizontally to the right in **Figure 1A**, and in the reverse case we move straight upwards. Those paths resemble conversions in a thermodynamic cycle, because $E$[ρ] is a state function. Thus, as a definition, a *path toward exactness* is a variation in ρ and $E$[ρ] that improves ρ and/or $E$[ρ] without deteriorating the other. In the above numerical definition of exactness, this path is well-defined by the errors of ρ and Δ$E$[ρ], which is an *observable* functional of ρ (which $E$[ρ] is not).

*Straying* from the path is then any change in a functional that increases error in ρ or Δ$E$[ρ]; thus the previously discussed straying[2] represents the dashed arrows on the diagonal from the top left toward the bottom right of **Figure 1A**. The *direct path* toward exactness is defined as the path that reduces errors in ρ or E[ρ] by the same proportion, represented by the diagonal lines moving from bottom right towards top left. It is an important special type of transformation of ρ → ρ$_{exact}$ and $E$[ρ] → $E_{exact}$[ρ$_{exact}$] where both transform by the same scale factor. One may improve functionals gradually toward the exact functional in generations with more mathematical terms of the gradient expansions, or in the numerical definition, by reducing errors, or fundamentally, by obeying more exact bounds (functionals marked " and ' in **Figure 1A**), or ideally several of these features in combinations, with some improving ρ more than the $E[ρ]$ or *vice versa*. These paths will deviate from the direct path but are not "straying" but rather converging toward exactness.

It is now useful to consider all real functionals as trial functionals of the exact functional. For the energy of any trial density ρ' studied by another functional $E[ρ']$[4],

$$E[ρ'] \geq E_0 \qquad (1)$$

where $E_0$ is the lowest possible energy (or "ground state" energy) $E[ρ]$ obtained from that functional with its variationally optimized ρ. Thus, any other ρ' will produce a larger $E[ρ']$. Because of equation (1), the native combination, e.g. $E'[ρ']$, is always lower in energy than any of the states vertically above or below it, e.g. $E'[ρ'']$ and even $E'[ρ_{exact}]$. By inspection of **Figure 1A**, the top right horizontal process corresponds to minus the functional-derived error, $-\Delta E_F' = E_{exact}(ρ_{exact}) - E'(ρ_{exact})$ for the functional $E'[ρ']$ in the definition by Burke and co-workers[11], whereas the left vertical process of the upper right cycle corresponds to minus the density-derived error, $-\Delta E_D' = E'(ρ_{exact}) - E'(ρ') \geq 0$, where the last inequality follows from the variational Hohenberg-Kohn principle, with equal being the case only for ρ' = $ρ_{exact}$. Burke and co-workers already discussed that a majority of systems behave normally (at least beyond Thomas-Fermi theory) in the sense that $\Delta E_D' \ll \Delta E_F'$.

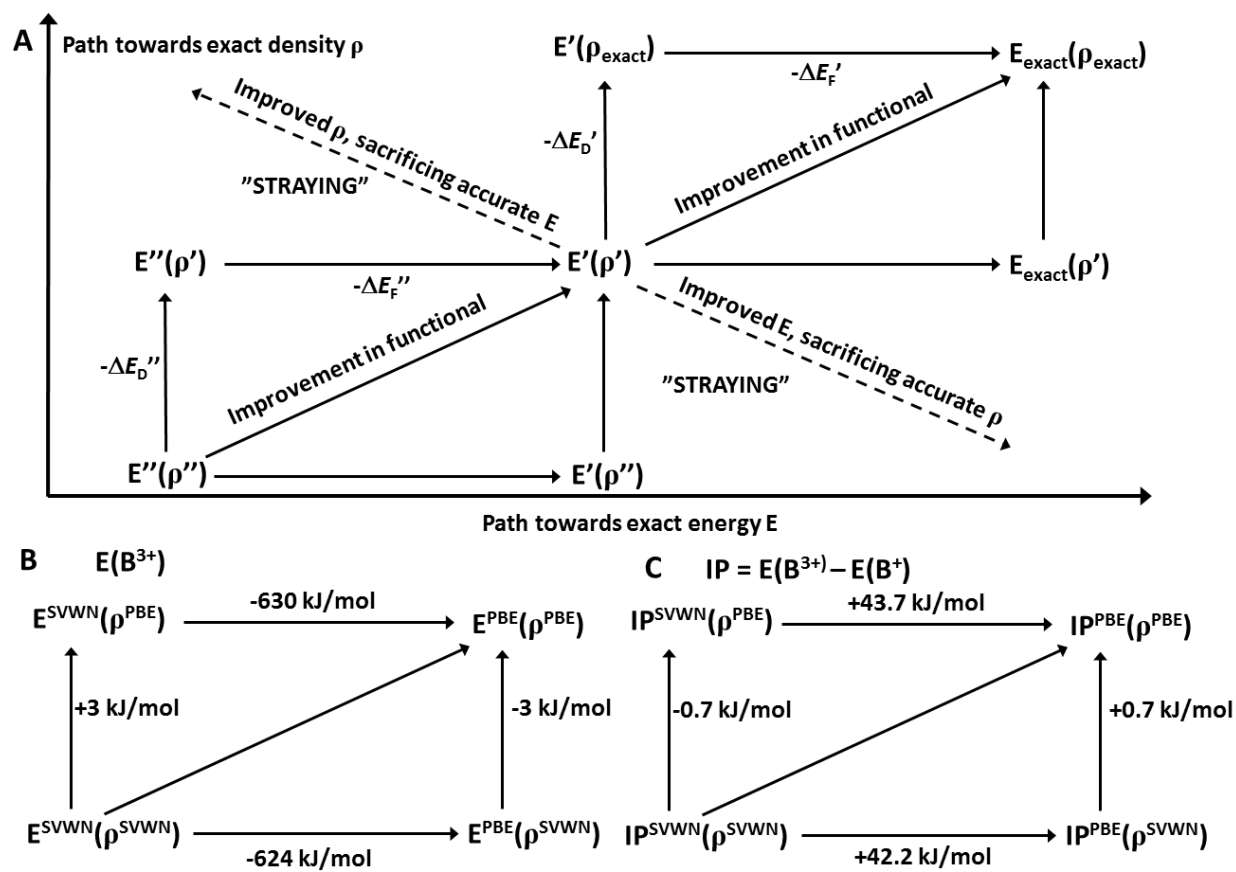

**Figure 1.** Simplified scheme of the paths toward more accurate density functionals.

The present work uses the Hohenberg-Kohn trial density concept and the fact that $E[\rho]$ is a state function to estimate the chemical significance of errors in $\rho$ as an alternative to using normalized relative errors that may not show the energy impact directly[2]. The definitions of paths toward or away from exactness in Figure 1 may help in defining norms for calculation. To do so, we can evaluate in a computationally cheap way (without using costly correlated wave function methods) the chemical significance of errors in $\rho$ that are intensively discussed[2]. More specifically, categorizing electronic systems and functionals as *practically normal* or *abnormal* based on the sensitivity of derived energies to variations in *trial* densities (rather than errors vs. the elusive exact density) turns out to be useful since exact densities are generally computationally inaccessible for most electronic systems, including the most interesting targets for development of new functionals. To demonstrate the protocol, it is first applied to the previously studied compact ions[2], and subsequently to the more chemically relevant first ionization potentials of atoms and the bond dissociation energies of molecules. It is shown that "practical normality" is a good proxy for "exact normality" and all these electronic systems are "practically normal", i.e. even large variations in densities produce insignificant energy variations, and thus the systems are *not* good norms for estimating progress on the path defined in **Figure 1**.

**Methods.**

All computations were performed using the Turbomole software, version 7.0[13] and the resolution of identify approximation was used to speed up all calculations[14,15]. For the native combinations of methods (i.e. those using their own converged $\rho$), all densities and energies were converged to $10^{-7}$ a.u. The energies of all molecules, ions, and atoms were computed using the aug-cc-pV5Z basis set[16] (numerical data in Table S1), except the atomic ions $B^+$, $B^{3+}$, $C^{2+}$, $C^{4+}$, $N^{3+}$, $N^{5+}$, $O^{4+}$, $O^{6+}$, $F^{5+}$, $F^{7+}$, $Ne^{6+}$, and $Ne^{8+}$, which were studied with the aug-cc-pωCV5Z basis set for direct comparison to previous studies using this basis set[2][10]. To investigate the effect of basis set on $\rho$ variations and $\Delta E_D$', the def-TZVP basis set[17] was also used for some systems.

Ionization potentials (IP) were computed as:

$$IP(X) = E_{el}(X^+) - E_{el}(X) \qquad (2)$$

where $E_{el}$ represents the electronic energy obtained with a method using a specified frozen density obtained by a converged previous computation either by the same method or another method. The double-ionization potentials (not to be confused with the second ionization potentials) studied for $B^+$, $C^{2+}$, $N^{3+}$, $O^{4+}$, $F^{5+}$, and $Ne^{6+}$ were computed as described previously[10]:

$$\text{IP}(X^{n+}) = E_{el}(X^{(n+2)+}) - E_{el}(X^{n+}) \tag{3}$$

The advantage of these experimentally available energies (they can be derived from ionization potentials of various order[10]) is that they involve strict comparison of the quality of $E[\rho]$ of the same studied closed-shell $1s^2$ and $1s^22s^2$ configurations without a need to invoke additional open-shell configurations, as would be the case for first ionization potentials.

The bond dissociation energies (*BDE*) were computed as:

$$BDE(XY) = E_{el}(X) + E_{el}(Y) - E_{el}(XY) - ZPE(XY) \tag{4}$$

where $E_{el}(X)$ is the electronic energy of species X computed and $ZPE(XY)$ is the vibrational zero-point energy of XY. In some cases, only the energies without *ZPE* were compared as *ZPE* does not affect the study of $\rho$ because *ZPE* is a constant of the geometry optimized with BP86/def2-TZVPP and generally changes only 1-2 kJ/mol with method[18,19]. The geometry used to compute the Co-C BDE was the complete, large cobalamin model including side chains that was previously published[20]. For def2-TZVPP this involves 4538 basis functions for the calculation of the energies of the complex and 338 closed-shell doubly occupied MOs (3710 basis functions for the cofactor radical where the 5'-deoxyadenosyl group has dissociated). The BDEs were in these cases computed using the conductor-like screening model (Cosmo)[21] as implemented in Turbomole[22] with ε = 80 and the dispersion D3 correction term by Grimme[23] except for SVWN, to produce a more realistic estimate of the true Co-C BDE, which depends on solvation and dispersion effects[20] (which the diatomic data do not since they are known experimentally in vacuum and dispersion effects are < 1 kJ/mol on the BDEs of these simple diatomic molecules[24]).

To identify "practically normal" systems and thereby estimate the significance of errors and deviations from the path in **Figure 1**, the recipe is to choose a small but diverse trial set of e.g. four functionals, ideally from four different rungs in Jacob's ladder[25]. It is assumed that the average variations in this set are similar to the typical absolute error in ρ. If local functionals such as SVWN are included, this is probably correct (assuming that Jacob's ladder implies increased exactness). To illustrate the approach, four diverse density functionals were chosen for the main trial set: B3LYP, PBE, SVWN, and M06. B3LYP[26–28] represents a lightly parameterized hybrid GGA functional with 25% HF exchange, and M06 represents a heavily parameterized meta hybrid with 27% HF exchange [29], PBE is a GGA non-hybrid[30], and SVWN is composed of the Slater exchange functional and the local VWN correlation functional[31] and thus represents the Local Spin Density Approximation (LSDA). These functionals ranked very differently in terms of the accuracy for ρ of small atomic ions, with the largest maximum normalized error in ρ seen for SVWN (3.725), with the other three spreading conveniently at 2.495 (PBE), 2.123 (B3LYP), and 1.838 (M06)[2]. Accordingly, both numerically and algorithmically they represent four very distinct functionals as desired to estimate fairly the sensitivity of E towards ρ-variations. Subsequently, the results were validated by using an even more diverse (in terms of ρ-variations) trial test set consisting of M06-2X (1.463), PBE (2.495), SVWN (3.725), and PBE0 (1.675), with maximum errors in ρ reported in parenthesis from previous work[2].

In order to study numerically the effect of ρ vs. $E[\rho]$, the trial density concept was used in the following way: In addition to all "native" computations $E[\rho]$, computations were also done for *all* combinations of $E[\rho']$ and $E'[\rho]$ for the trial test set, where $E$ and $E'$ signifies energies obtained from two different functionals using fixed densities obtained by previous convergence with the other functional. For a trial set of $N$ functionals, this implies $N^2$ energy calculations for each electronic system. Fixed densities were obtained by allowing only a single energy evaluation of the new functional on the converged density and having infinite thresholds for convergence of the density ($denconv in Turbomole). The densities along an axis (x) of the coordinate system of the

atom or atomic ion were in all cases printed to text files and it was confirmed that densities were fixed and identical after calculation when used as trial densities.

**Results and Discussion.**

**Using Trial Densities To Estimate ρ-Derived Errors.** The formally defined error[11] in a functional due to its density, $\Delta E_D'$ (upper vertical process in **Figure 1A**) is not known unless $\rho_{exact}$ and $E_{exact}[\rho_{exact}]$ is known, which is only approximately the case for very small electronic systems that can be computed by correlated methods such as CCSD(T) with extensive basis sets, and even this method will fail for more complex electronic systems that are often interesting targets for developing new functionals. Because of this obstacle, we cannot systematically improve functionals along both axes of **Figure 1A** using $\Delta E_D'$ directly. Instead, I define a modified version of the normality proposed by Burke, which was based on the full error vs. the exact functional. To do so I use the same definition applied to the *left* thermodynamic cycle in **Figure 1A**. In any study of a non-exact functional $E''[\rho'']$ (which will always be the case), we can compute exactly the corresponding value $E''[\rho'] - E''[\rho'']$ where $\rho'$ is a trial density. The computed value $\Delta E_D''$ is the trial density analog of $\Delta E_D'$ and equals $\Delta E_D'$ in the case where $\rho' = \rho_{exact}$. I refer to this number as $\Delta E_D''$ in this paper to signify that it was estimated from trial densities. *I define "practically normal" systems as those for which a diverse trial set of functionals produce an average $\Delta E_D'' < 4$ kJ/mol (chemical accuracy).*

The differences in ρ obtained for different pairs of the four functionals for 12 of the electronic systems studied by Medvedev et al.[2] are shown in Supporting Information, **Figures S1-S12** along the x-axis in units of bohr. The main differences between densities produced by these four diverse functionals are located near the nucleus ($x = 0$). Notably, PBE and B3LYP produce the most similar ρ, as shown by the yellow lines, but start to deviate at large $Z$ (compact density limit). The ρ produced by the local spin density approximation SVWN deviates the most from the other functionals (red, blue, and gray colors) for all 12 systems. However, M06 is also quite distinct from B3LYP and PBE in most cases, and interestingly, the Δρ of B3LYP-M06 and PBE-M06 displays non-monotonic behavior near the nucleus (i.e. the difference is largest at the nucleus but then has a second maximum deviation at +/- 0.1 bohr. This second maximum is reduced as $Z$ increases. The deviation in ρ(SVWN) grows monotonically with $Z$ as ρ becomes more compact and the lack of gradient manifests around the nucleus (the nuclear cusp condition implies that the change in ρ at the nucleus scales linearly with $Z$)[32]. Accordingly, in general, all errors in densities for the $1s^22s^2$ systems that dominate the total error in the previous study scale almost linearly with $Z$[10]. The difference between M06 and B3LYP/PBE is generally larger for $1s^2$ systems than for $1s^22s^2$ systems.

**Figure 1B** shows a numerical example for the $B^{3+}$ ion, which is one of the cations studied previously[2], using PBE and SVWN. For this system, as well as other systems (see numerical data in Supporting Information), while the absolute differences in $E_{el}$ are large and not meaningful on the non-variational "functional axis", the numbers along the vertical variational axis corresponding to $\Delta E_D$" are ~3 kJ/mol. Due to the variational principle applied to densities[4], the native combination produces smaller absolute energies. This is exemplified in **Figure 1B** from the signs +3 and -3 kJ/mol when computing energies according to a thermodynamic cycle, where the energy of change equals the energy of the end state minus the energy of the start state. Despite the major differences in these functionals, the error is already small in the absolute (non-observable) $E_{el}$.

For the ions studied previously[2], the quality of E[ρ] can be estimated by computing the energy of removing the two 2s electrons, e.g. $E(B^{3+}) – E(B^+)$, which can be compared directly to accurate experimental data from NIST[10]. As seen in **Figure 1C**, for this energy, errors using a trial density reduce in both directions to 0.7 kJ/mol. The values of these energies are many tens of eV, as discussed previously[10]. In the boron case, the experimental number is 63.1 eV. The ρ-derived error using SVWN as trial density instead for PBE is 0.7 kJ/mol or about 0.01% of the total computed number, and ~4% of the total error that PBE makes (18.3 kJ/mol error). The four values are SVWN(SVWN) = 6025.7 kJ/mol, SVWN(PBE) = 6025.0 kJ/mol, PBE(PBE) = 6068.7 kJ/mol, and PBE(SVWN) = 6067.9 kJ/mol (the experimental number is 6086.9 kJ/mol).

Since the difference between the experimental value and PBE is only a third of the difference between experiment and SVWN, and considering the ρ improvements using PBE vs. SVWN (SI, Figure S1, S2 and previous work[2]), we could use the jargon[2] that PBE is more "exact" for $B^+$ and $B^{3+}$ than SVWN, and roughly speaking two-thirds of the way towards "exactness" from SVWN for this system. From these distance-metric considerations, we can argue that $\Delta E_D'$ will continue to be very small compared to $\Delta E_F'$ also in the final step towards exactness, although we do not know the exact $E[\rho]$ for the boron ions (or any other system).

The results in **Figure 1C** suggest that trial densities such as PBE may sometimes be used to search for exact functionals, because the error made in ρ is quite small, consistent with previous suggestions[11]. Systems such as $B^+$ and $B^{3+}$ will have $\Delta E_D'$ so small that it becomes insignificant, i.e. any optimization of ρ reduces $\Delta E_D'$ from 1 kJ/mol to a smaller number within the uncertainty of other approximations of the non-relativistic Born-Oppenheimer-type Kohn-Sham calculation.

Because $\Delta E_D'$ of a calculation depends both on system and functional, functionals may also be defined as normal or abnormal. I define here "practically normal" as a functional *and* a system (external potential) that *together* give $\Delta E_D'$ < 4 kJ/mol (chemical accuracy). Thus, *if a calculation gives $\Delta E_D'$ < 4 kJ/mol for a suitable diverse test set of functionals, the system and the functional are normal*. Again, *we can search for the exact functional for normal systems using fixed trial densities to within chemical accuracy*, making the approximation $\Delta E_D' \sim \Delta E_D''$. In contrast, *universality* will only be effectively approached by studying abnormal systems where the exchange-correlation functional changes more substantially with variations in ρ[11].

Similar computations were done for all combinations of the trial test set SVWN, B3LYP, PBE, and M06, with all the cations $B^+$, $B^{3+}$, $C^{2+}$, $C^{4+}$, $N^{3+}$, $N^{5+}$, $O^{4+}$, $O^{6+}$, $F^{5+}$, $F^{7+}$, $Ne^{6+}$, and $Ne^{8+}$, as studied previously[2,10] (see Supporting Information **Tables S1-S6** for numerical details). To make data more accessible, the average of the deviations in double IPs caused by the tree other trial densities were computed for all six cases; these average values represent sensitivities of the computed observable energy to the use of trial densities and are shown for all four functionals in **Figure 2A** (for B3LYP), **Figure 2B** (for PBE), **Figure 2C** (for SVWN), and **Figure 2D** (for M06).

From **Figure 2**, it can be seen that the effect of using a trial density on the numerically very large energy of removing two 2s electrons (43 – 446 eV experimentally) is maximally 3 kJ/mol. M06 produces the "worst" trial densities for B3LYP and SVWN (**Figure 2A, C**), and that B3LYP and PBE provide uniformly accurate trial densities for each other. In fact, instead of computing the more expensive $E_{B3LYP}(\rho_{B3LYP})$, one may compute $E_{B3LYP}(\rho_{PBE})$ by a single energy calculation on the frozen, previously optimized (without expensive Hartree-Fock exchange) $\rho_{PBE}$ and obtain results for these large numbers that differ from the fully optimized B3LYP values by < ½ kJ/mol *in all six cases* (**Figure 2A**). This confirms the suggestion[10] that the errors in ρ of these systems[2] are not numerically relevant and thus "normal" in the terminology of Burke and co-workers[11].

The density differences plotted in **Figures S1-S12** agree well with the energetics of **Figure 2**; thus, B3LYP/PBE are the most similar behaving functionals and can be used with high confidence as trial densities for each other for these systems. Interestingly, these differences provide a semi-quantitative metric of the similarity of functionals that is consistent both in ρ and $E[\rho]$ space and thus may in principle be used as a quantification of degree of exactness, if a similar metric was applied to the exact functional, which is of course unknown.

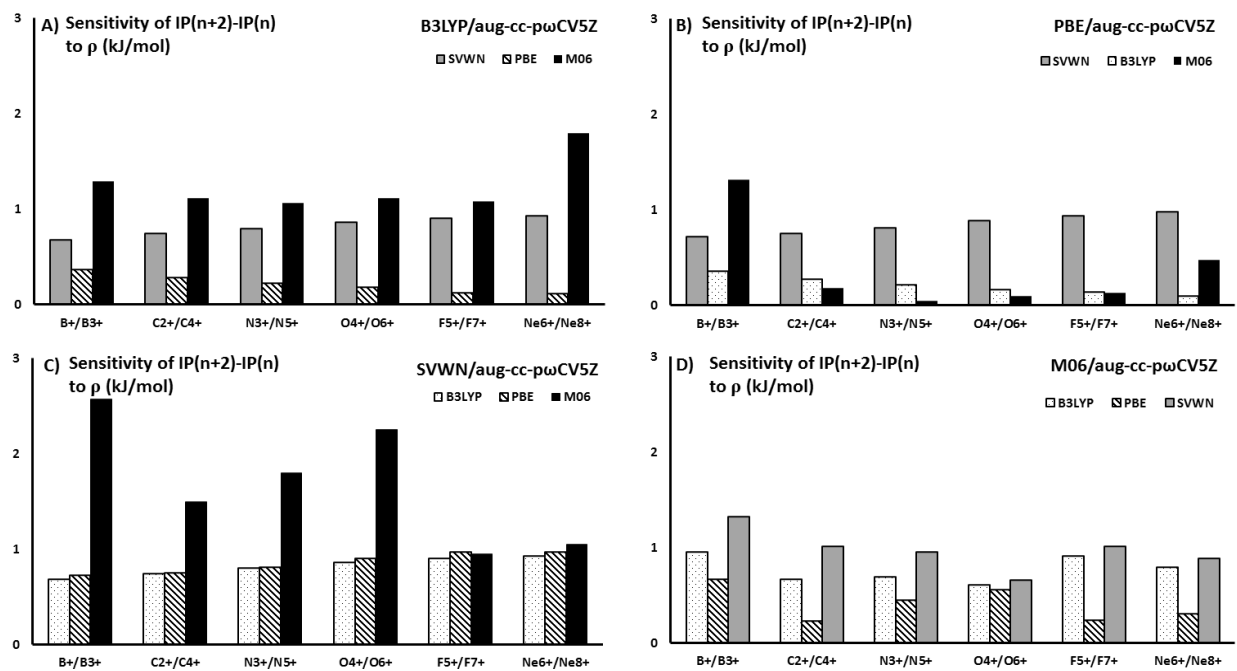

**Figure 2.** Mean absolute deviation between double IPs computed with the native E[ρ] and using trial densities E[ρ']. **A)** B3LYP energies using SVWN, PBE, and M06 for ρ'. **B)** PBE energies using SVWN, B3LYP, and M06 for ρ'. **C)** SVWN energies using B3LYP, PBE, and M06 for ρ'. **D)** M06 energies using SVWN, B3LYP, and PBE for ρ'.

To confirm that the trial set is estimates $\Delta E_D''$ well, the average $\Delta E_D''$ for the double IPs of $B^+$, $B^{3+}$, $C^{2+}$, $C^{4+}$, $N^{3+}$, $N^{5+}$, $O^{4+}$, $O^{6+}$, $F^{5+}$, $F^{7+}$, $Ne^{6+}$, and $Ne^{8+}$ were also compared with another trial set involving the more diverse (in terms of previously reported errors in densities) functionals M06-2X, PBE0, PBE, and SVWN. This set represents largest differences in errors of the densities and are compared to the M06-PBE-SVWN-B3LYP trial test set in **Figure S13**. As can be seen, the average energy effects of interchanging these densities remain < 2 kJ/mol, confirming the conclusion above.

The effects described here arise from the differential behavior of the exchange correlation potentials and are exemplified by PBE applied to $B^+$ and $Ne^{6+}$ in **Figure 3A-B** and **Figure C-D**, respectively. When the exchange correlation potential of the native density (**Figure 3A** and **Figure 3C**) is evaluated using a trial density, the differences in densities translate into an effect on the functional's exchange-correlation potential (**Figure 3B** and **Figure 3D**), which then translate into energy. It can be seen that whereas the absolute potentials are of course localized near the nuclei, the differential effects central to the discussion of the ρ-variations spread over much of the electronic system, explaining why these energy effects do not increase monotonically with nuclear charge and compactness of the electron density as the errors in ρ themselves do for these systems.

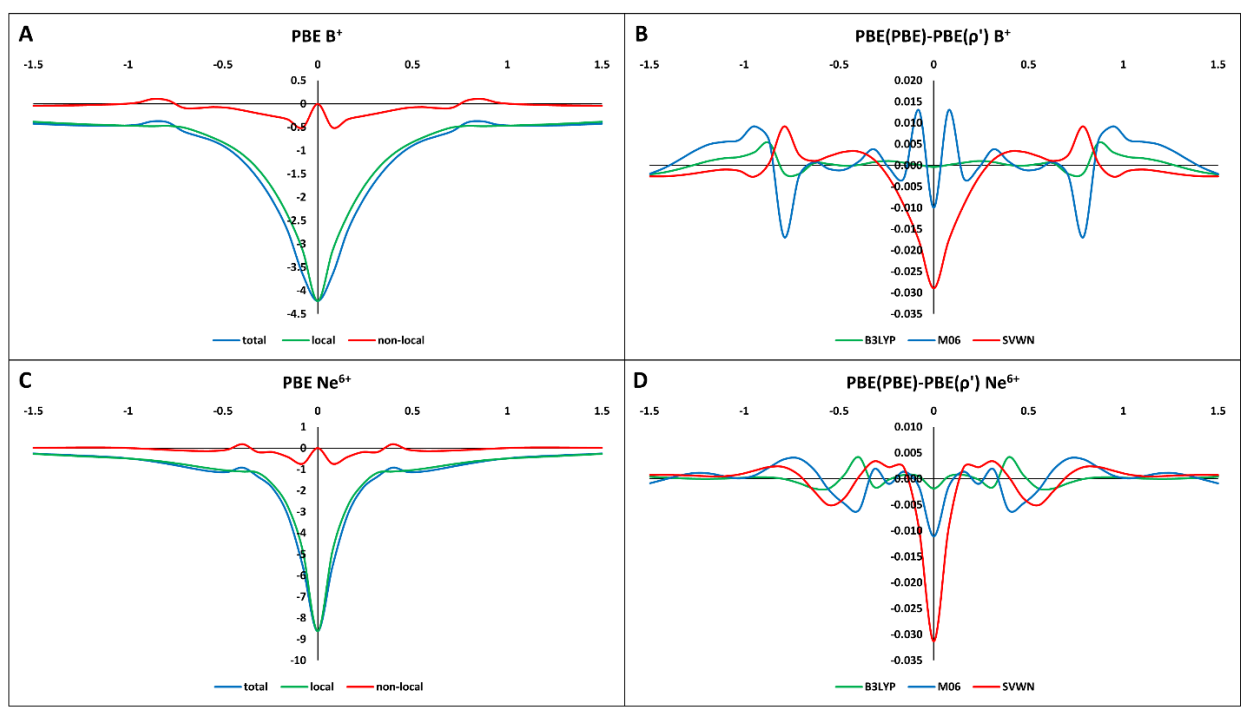

**Figure 3. A)** PBE exchange-correlation potential for $B^+$ applied to the native PBE density; **B)** same but applied to three other trial densities converged with other functionals; **C)** PBE exchange-correlation potential for $Ne^{6+}$ applied to the native PBE density; **D)** same but applied to three other trial densities converged with other functionals.

**First Ionization Potentials of Neutral Atoms.** In the following, the above procedure is applied to the study of the more chemically relevant first IPs of atoms. To avoid the complication of relativistic effects, only the first and second period of the periodic table were studied. The experimental data are in **Table S7**; Tables **S8-S10** show numerical details for native method combinations, whereas **Tables S11-S14** show results using trial densities. **Table 1** shows the computed IPs using the aug-cc-pV5Z basis set and the methods CCSD(T), CCSD, B3LYP, SVWN, PBE, and M06. For these experimentally observable energies, CCSD(T) achieves a mean absolute error (MAE) of ~1.5 kJ/mol, i.e. the description of the systems is essentially correct to within this limit (scalar relativistic effects are < 1 kJ/mol for these energies[10]). Among the four functionals, M06 displays the smallest MAE, probably because it was parameterized to fit these data[29]. The main question of interest here is how much $\rho$ contributes to the energy errors, because this should be required of a good density norm.

**Table 1. Errors in computed ionization potentials (aug-cc-pV5Z basis, in kJ/mol).**

|     | CCSD(T) | CCSD  | B3LYP  | PBE    | SVWN   | M06    |
| --- | ------- | ----- | ------ | ------ | ------ | ------ |
| H   | 0.49    | 0.49  | -1.99  | 0.46   | -55.50 | 1.00   |
| He  | -1.19   | -1.19 | 24.93  | -11.94 | -27.75 | 18.22  |
| Li  | -1.15   | -1.23 | 15.14  | 18.69  | 7.81   | -8.24  |
| Be  | -1.41   | -2.81 | -28.58 | -31.58 | -28.70 | -37.39 |
| B   | -1.75   | -5.12 | 34.62  | 36.27  | 33.72  | 13.09  |
| C   | -1.04   | -3.80 | 18.81  | 26.93  | 41.59  | 5.72   |
| N   | 0.04    | -2.70 | 3.25   | 18.21  | 45.07  | 19.40  |
| O   | -4.75   | -10.08 | 39.76 | 42.58  | 37.02  | 18.43  |
| F   | -2.58   | -9.11 | 18.65  | 21.95  | 51.31  | 11.87  |
| Ne  | 1.15    | -6.32 | 3.29   | 8.28   | 59.54  | 9.95   |
| *MAE* | *1.55* | *4.28* | *18.90* | *21.69* | *38.80* | *14.33* |

**Figure 4A and 4B** show the differential electron densities for two of these more electron-rich systems, Ne, and Ne$^+$. It is notable that the same patterns in the differential densities observed for the closed-shell s-configurations still hold for the more electron rich p-configurations, including the open-shell systems, exemplified by Ne$^+$ (**Figure 4B**).

**Figure 4C** and **4D** show the average sensitivity (mean absolute deviation between energies obtained using the native density and trial densities) using the aug-cc-pV5Z and def-TZVP basis sets, respectively (**Figure S14** shows the full plots without averaging. The data with def-TZVP are in **Tables S15 and S16**). The errors obtained with each native functional and density combination are given in **Table S17** for both basis sets. As seen from **Figures 4C** and **4D**, with both basis sets, the average effect of the trial densities on computed IPs is smaller than 2 kJ/mol except for Ne, where M06 deviates on average by 4 kJ/mol, more than any other functional including SVWN, although its densities are more similar to those of B3LYP and PBE. This relates to the variation of the exchange-correlation energy with density, i.e. to the exchange correlation potential functional, $v_{xc} = \partial E/\partial \rho$, e.g. **Figure 3B**. The meta feature and its heavy parameterization of M06 could affect the smoothness of the electron density near the nucleus where the energy and density effects are most important. This is a general tendency despite the distinct s- and p-electronic configurations and number of unpaired electrons, and regardless of basis set used (hydrogen and beryllium are exceptions). It is also interesting to observe that M06 is approximately doubly as basis-set sensitive as B3LYP (1.89 vs. 0.99 kJ/mol average, **Table S18**, for def-TZVP vs. aug-cc-pV5Z) and also more than doubly as sensitive to ρ as B3LYP (1.01 vs. 0.37 kJ/mol). In order words, it suggests that basis set and density sensitivity are correlated. The basis set sensitivity and the density sensitivity probably both arise from the same underlying behavior of the exchange correlation potential (**Figures 3B, 3D**), and e.g. smoothness constraints may improve M06 in this regard.

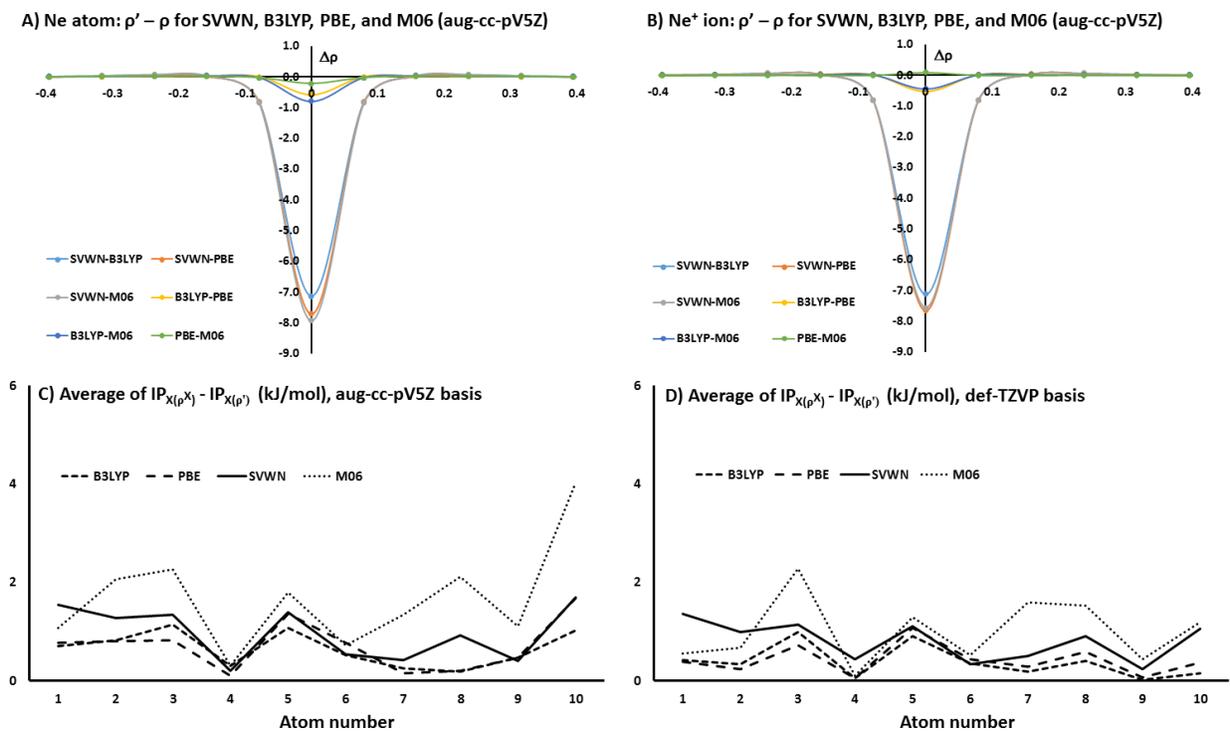

**Figure 4. A)** Differential electron densities ρ-ρ' for Ne atom; **B)** same for Ne⁺ ion; **C)** average sensitivity of four functionals to the use of trial ρ' from other functionals at the aug-cc-pV5Z basis set level; **D)** same at the def-TZVP basis set level.

Another relevant observation is an oscillating behavior in the sensitivity toward ρ', which reflects variable (practical) normality of the electronic systems; thus, $Z = 1, 4, 7$, and 9 show the highest "normality", whereas $Z = 3, 5, 8$, and 10 are more "abnormal". These differences can be traced to the difficulty of describing ρ when electronic configurations change qualitatively upon ionization (simply speaking, in terms of their n, l, and s quantum numbers). Thus, Li goes from the high-energy delocalized 2s to a fully symmetric closed-shell $1s^2$ configuration in Li⁺. B goes from an open-shell $2p^1$ configuration to a fully symmetric closed-shell $1s^2 2s^2$ B⁺ state. O has to remove a paired p-electron from $[He]2s^2 2p^4$ to produce a particularly stable maximum-spin, half-occupied p-shell (for N, p-orbital pairing is avoided; in F, there are two paired p-orbitals so the effect on pairing is relatively smaller). Ne goes from a completely filled closed shell to a delocalized open-shell hole in the 2p-shell. Regardless of this, all these systems are practically normal if we use chemical accuracy (4 kJ/mol) as a threshold.

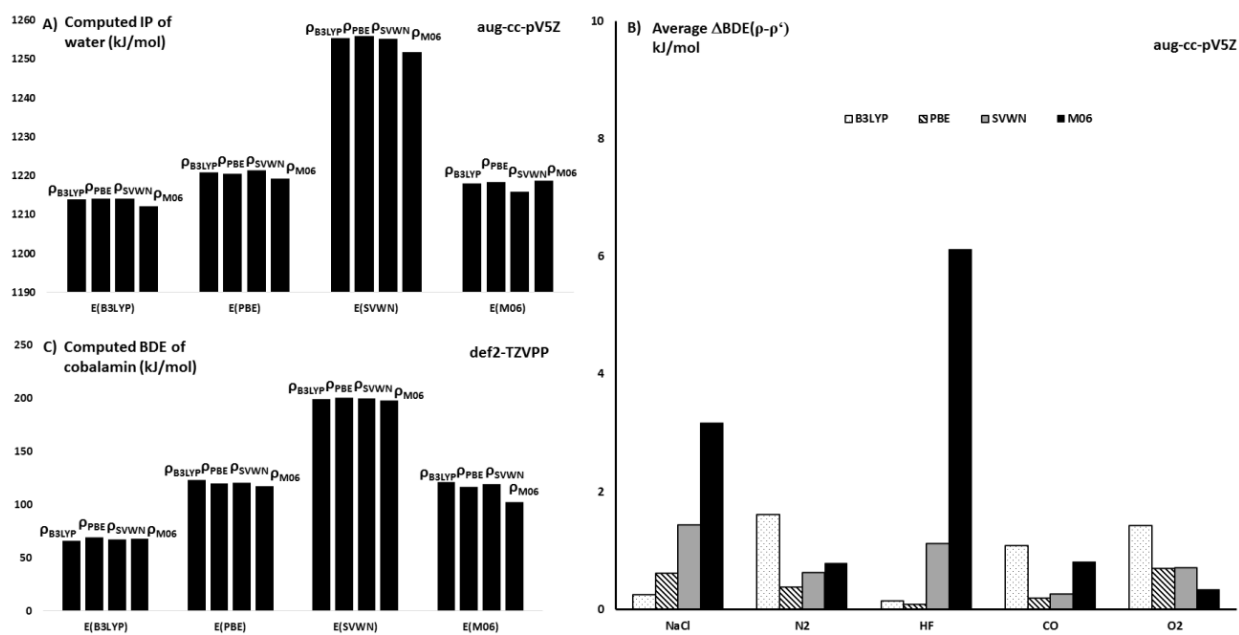

**Figure 5. A)** Computed ionization potentials (aug-cc-pV5Z) of H$_2$O using all 16 trial combinations of B3LYP, PBE, SVWN, and M06. **B)** Average change in computed BDE of NaCl, N$_2$, HF, CO, and O$_2$ when using other densities than the native density, for B3LYP, PBE, SVWN, and M06 and the aug-cc-pV5Z basis set. **C)** Computed Co-C BDE of 5'-deoxyadenosylcobalamin at the def2-TZVPP basis set level, using all 16 possible trial combinations.

**Examples for Molecular Systems.** To illustrate the procedure on molecular systems, three more cases were studied: The ionization potential of the water molecule, computed BDE of diatomic molecules, and the Co-C BDE of the large coenzyme 5'-deoxyadenosylcobalamin, one of the active forms of vitamin B$_{12}$. The data set of diatomic molecules represents ionic bonding (NaCl, HF), covalent bonding (N$_2$, O$_2$), and polar covalent bonding (CO) as well as single (HF, NaCl), double (O$_2$), and triple bond (N$_2$, CO) character. The variations of these computed observables with changes in ρ' are summarized in **Figure 5**.

**Figure 5A** shows the IP computed for the water molecule using any of the 16 combinations of density functionals and densities. The computed electronic energies are described in detail in the Supporting Information **Tables S19-S23**. The experimental IP is 1220.6 kJ/mol. The computed values using B3LYP, PBE, SVWN, and M06 are 1213.8, 1220.5, 1255.2, and 1218.8 kJ/mol, respectively. Computing the IP using trial densities produce similar IP values to within 4 kJ/mol; this largest variation is seen for the local SVWN functional. Again, the PBE and B3LYP functionals give results within 1 kJ/mol when using the other as trial density, and M06 is somewhat more sensitive than these two to the nature of ρ'.

**Figure 5B** show the average change in the BDE of NaCl, N$_2$, HF, CO, and O$_2$ for each functional using its own functional vs. using the other densities as trial densities, referred to as ΔBDE(ρ-ρ'), all computed with the aug-cc-pV5Z basis set for the five diatomic molecules. A small ΔBDE(ρ-ρ') implies that the functional is not very sensitive to the use of trial density for the particular electronic system of interest. The BDEs for the five molecules were also computed with CCSD(T)/aug-cc-pV5Z and CCSD/aug-cc-pV5Z including ZPE (**Table S24**) to estimate accuracy of the densities produced from this basis set. A mean absolute error (MAE) for CCSD(T) of 5.3 kJ/mol is obtained using the aug-cc-pV5Z basis set (**Table S25**). Part of the error in CCSD(T) relates to relativistic corrections (of the order of ~1 kJ/mol, estimated from scalar-relativistic contributions, and enthalpy corrections (numerical results in **Tables S26**-**S27** for native methods and in **Tables S28-S31** using trial densities).

For these systems, B3LYP and PBE produce small variations in BDE when ρ' is varied. It confirms that when studying also "normal" molecular systems, one can use *any* reasonably accurate ρ and save substantial computer time. However, in the two cases of strong ionic bonding (NaCl and HF) $\rho_{M06}$ displays an average sensitivity of ~6 kJ/mol, even though the electronic systems are practically normal. Again this reflects that M06 deviates the most from the other functionals near the energy-wise important nuclei. As seen, "normality" of the systems is not directly related to the strength of the bonds (the dense electron limit) i.e. *all these systems are practically normal despite their distinct bonding behaviors*, yet are important benchmarks for describing accurate thermochemistry. Abnormal systems may be diffuse or delocalized densities, e.g. anions or dissociating states, as discussed previously[11]. The use of orbital-dependent HF densities as "trial" densities, $\Delta E[\rho']$ where $\rho' = \rho_{HF}$, is often much more accurate for such abnormal systems[33], where the self-interaction error also manifests.

As a final example of a large system, **Figure 5C** shows the computed Co-C BDEs for 5'-deoxyadenosylcobalamin, the active form of vitamin $B_{12}$, with 676 electrons. For such large systems, accurate thermochemistry depends critically on the inclusion of dispersion corrections and solvent effects[20]; the Cosmo model[21] for water was used as well as D3 corrections[23], except for SVWN. As can be seen from **Figure 5C** (numerical data in **Table S32**), the Co-C BDEs are generally not very sensitive to the used $\rho'$ even for this large, complex electronic system. As described previously, for this particular system, PBE gives more accurate results (the experimental value is ~130 kJ/mol)[20]. It is surprising that a good estimate is obtained regardless of the $\rho'$ used with PBE, implying that even a challenging problem like the Co-C BDE of vitamin $B_{12}$ belong to the class of "practically normal" systems. Consequently, we can obtain the Co-C BDE of any normal functional very quickly to using any normal, previously converged density such as $\rho_{PBE}$, making errors < 4 kJ/mol (chemical accuracy); the single-point energy evaluations on fixed densities are multi-fold faster than a fully converged computation. For example, the total cpu time for converging the full density and energy for the full system was 17h 51m with B3LYP, but only 17 minutes for the subsequent PBE calculation. The full PBE convergence took 3h 39m, and the B3LYP calculation using this density 4h 27m (very similar, 4h 31m if using the M06 density). Thus in particular for non-hybrid functionals that do not require evaluation of the exchange integrals, the saved computer time is a factor of ten for this system, losing an accuracy of 3-4 kJ/mol. Consequently, even though B3LYP fails substantially in describing this BDE, as discussed in the literature[34,35], PBE performs very well even on the B3LYP density; in other words, in the "thermodynamic cycle" of PBE and B3LYP, the change in ρ affects the energy 16 times less than the change in functional. M06 stands out by having the largest sensitivity to trial density by a factor of 2 (8.3 kJ/mol vs. 1.2-3.9 kJ/mol for the other three, **Table S32**), again confirming its more sensitive exchange correlation potential vs. the other functionals, a feature that also seems to make it more basis-set dependent.

**Conclusions.**

In the search for more exact and generally applicable density functionals, the lack of systematic recipes of improvement and of well-defined "paths" toward exactness is a major challenge. As discussed in this work, the Hohenberg-Kohn theorems are central to this challenge. Specifically, dividing errors in a functional into a term for ρ ($\Delta E_D'$) and one for the effect of the functional on ρ ($\Delta E_F'$) (including manipulations of its gradient, Laplacian, etc.), as recently suggested by Burke and co-workers[11], enables a formal strategy for improving functionals, as outlined in **Figure 1**. Specifically, the *direct* path toward exactness from any functional is defined as the straight diagonal line towards the upper right in **Figure 1A**: This path is defined by $\Delta E_D' \to 0$ and $\Delta E_F' \to 0$ by the same scale factor of the total error in $E'[\rho']$ as $\rho' \to \rho_{exact}$ and $E'[\rho'] \to E_{exact}[\rho_{exact}]$. "Straying"[2] then becomes well-defined as any change in a functional that increases $\Delta E_D'$ *or* $\Delta E_F'$.

To determine deviation from the path, it is suggested to separate energy effects due to ρ and due to the way ρ is converted into a property of interest (here: $E[\rho]$) by the functional. The exact $\Delta E_D'$ suggested by Burke et al.[11] (i.e. compared to exact densities and energies) is generally inaccessible, preventing its use in estimating deviations from or progress on the path defined in **Figure 1A** for most electronic systems of interest. To solve this problem, I use trial densities ρ' for a suitably chosen trial test set of functionals that should i) be small (4-5 functionals) to make sensitivity analysis computationally tractable; ii) consist of diverse functionals, preferably from different rungs of Jacob's ladder, to estimate as broadly as possible the density sensitivity. The present analysis shows that systems and functionals can both be classified as "practically normal" or abnormal via their average $\Delta E_D''$ for the defined trial test set. I define here a practically normal calculation as one that has $\Delta E_D'' < 4$ kJ/mol (chemical accuracy).

Using the protocol and definitions, a range of electronic systems are studied, starting with the systems previously studied by Medvedev et al.[2] because they specifically suggested a norm for densities, but then moving to more chemically relevant systems. Additional conclusions arising from these calculations are:

1) Variations in E[ρ'] using a suitably diverse test set of trial densities offers a fast and efficient way to estimate the sensitivity of the energy to variations in ρ; this enables a fast way to estimate if variations in ρ have measurable effects on E[ρ], i.e. if they are useful norms for assessing functional performance or are within the noise of computational chemistry.

2) Previously studied[2] compact atomic ions used to conclude that density functionals are straying are "practically normal" in the sense that $\Delta E_D'' < 4$ kJ/mol. Thus, this straying is not numerically significant vs. other approximations in the Kohn-Sham procedure, such as the use of Gaussian basis functions for very compact electron densities, whose main differences manifest near the nuclei. For example, IPs produced by PBE and B3LYP using each other's densities are typically ~1 kJ/mol. The protocol thus rules out these systems as good norms.

3) More chemically relevant atoms and monocations for the first and second row ($Z$ = 1-10) studied in this work, and accordingly also first ionization energies, are also practically normal, and the energy consequences of varying densities are thus too small to be chemically relevant and mostly beyond assessment of accuracy considering other approximations done in the Kohn-Sham Born-Oppenheimer formalism that could affect energies up to a few kJ/mol.

4) Bond dissociation energies of diatomic molecules are practically normal for large basis sets, but can become less normal for large systems or smaller basis sets; thus while B3LYP and PBE give very similar results for the Co-C BDE of the large complex 5'-deoxyadenosylcobalamin if the other's $\rho'$ is used for calculation, M06 gives very different results depending on $\rho'$, probably because this functional has a more sensitive exchange correlation functional, viz. **Figure 4**. Still however, even large electronic systems such as the cobalamins can be studied with simpler trial densities to speed up single point computations by loss of only a few kJ/mol in accuracy (e.g. using B3LYP with PBE trial densities).

This analysis may be useful in focusing efforts towards developing new functionals. For example, $\rho'$(PBE) will serve excellently as trial density in the search for $\rho$ if the benchmarked electronic systems are normal (e.g. IPs and BDEs computed by the separation method[11]); but to move further, one needs to improve functionals for abnormal systems where errors in $\rho'$ become significant vs. chemical accuracy. For errors smaller than this, it is hardly testable whether this contributes to exactness or not, because the uncertainty resembles that of other approximations tacitly applied. Therefore, an absolute definition of exact normality as $\Delta E_D' < 4$ kJ/mol and practical normality as $\Delta E_D'' < 4$ kJ/mol (chemical accuracy) seems reasonable.

**Supplementary Information.** The supplementary information file contains figures of differential density plots for the four functionals for $B^+$, $B^{3+}$, $C^{2+}$, $C^{4+}$, $N^{3+}$, $N^{5+}$, $O^{4+}$, $O^{6+}$, $F^{5+}$, $F^{7+}$, $Ne^{6+}$, and $Ne^{8+}$ (Figures S1-S12); average $\Delta E_D$" for the double IPs of $B^+$, $B^{3+}$, $C^{2+}$, $C^{4+}$, $N^{3+}$, $N^{5+}$, $O^{4+}$, $O^{6+}$, $F^{5+}$, $F^{7+}$, $Ne^{6+}$, and $Ne^{8+}$ using two different trial tests sets (Figure S13); plots of the difference in IPs obtained for individual functionals when using trial densities (Figure S14); electronic energies and double IPs for $B^+$, $B^{3+}$, $C^{2+}$, $C^{4+}$, $N^{3+}$, $N^{5+}$, $O^{4+}$, $O^{6+}$, $F^{5+}$, $F^{7+}$, $Ne^{6+}$, and $Ne^{8+}$ (Tables S1-S6); experimental IPs (Table S7); electronic energies and IPs of neutral atoms computed with all methods at aug-cc-pV5Z (Tables S8-S14) and def-TZVP (Tables S15-S16); errors in computed IPs (Table S17); basis set sensitivities of IPs (Table S18); data for the water molecule (Tables S19-S23); ZPEs (Table S24) and computed BDEs (Table S25) of diatomic molecules; computed electronic energies and $D_0$ (without ZPE) of diatomic molecules (Tables S26-S31); computed Co-C BDE of 5'-deoxyadenosylcobalamin, and sensitivity to use of trial density (Table S32).

**Acknowledgements.** The Aarhus Supercomputer Center (formerly part of the Danish Center for Scientific Computing) is gratefully acknowledged for providing computer time for this project.

**SUPPORTING INFORMATION**

**Energy vs. density on paths toward exact density functionals**

Kasper P. Kepp*

*Technical University of Denmark, DTU Chemistry, Building 206, 2800 Kgs. Lyngby, DK – Denmark.* *Phone: +045 45 25 24 09. E-mail: kpj@kemi.dtu.dk

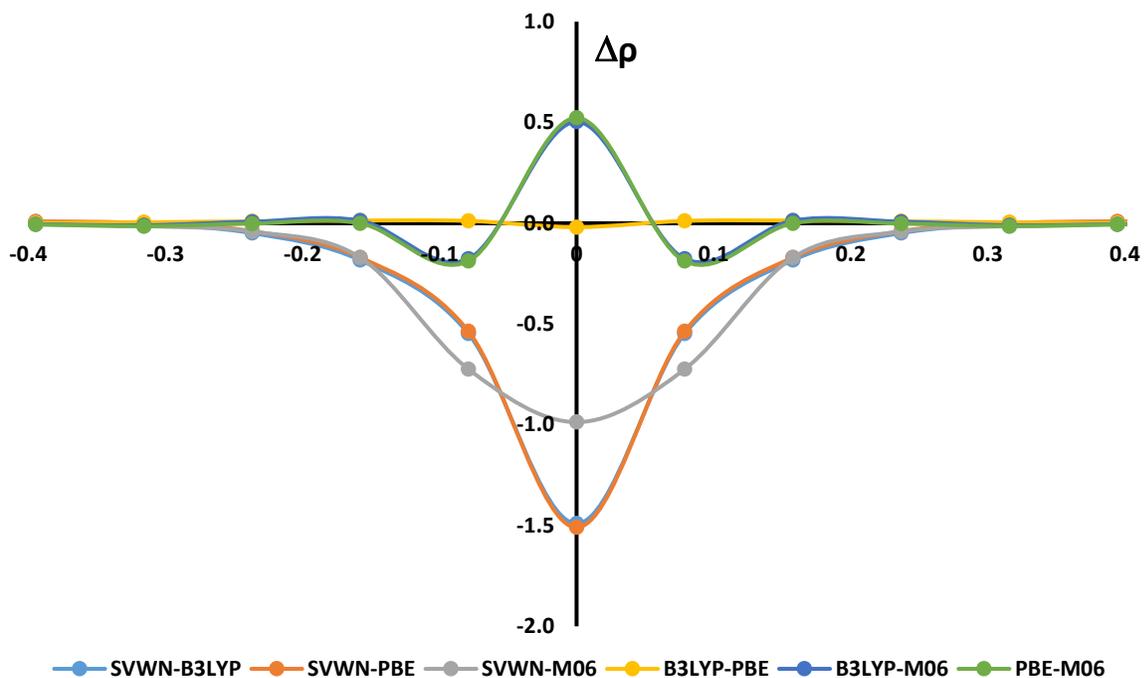

Figure S1. The six density difference plots of the four studied density functionals, B$^+$ ion. (aug-cc-pωCV5Z)

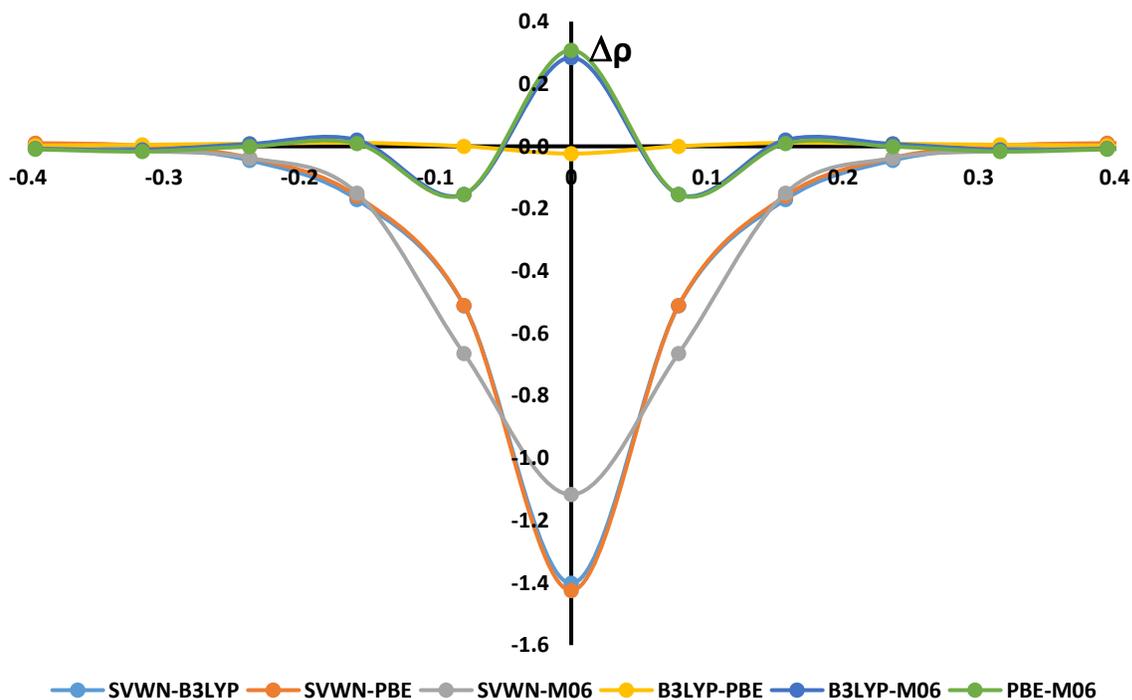

Figure S2. The six density difference plots of the four studied density functionals, B$^{3+}$ ion. (aug-cc-pωCV5Z)

**Figure S3.** The six density difference plots of the four studied density functionals, $C^{2+}$ ion. (aug-cc-pωCV5Z)

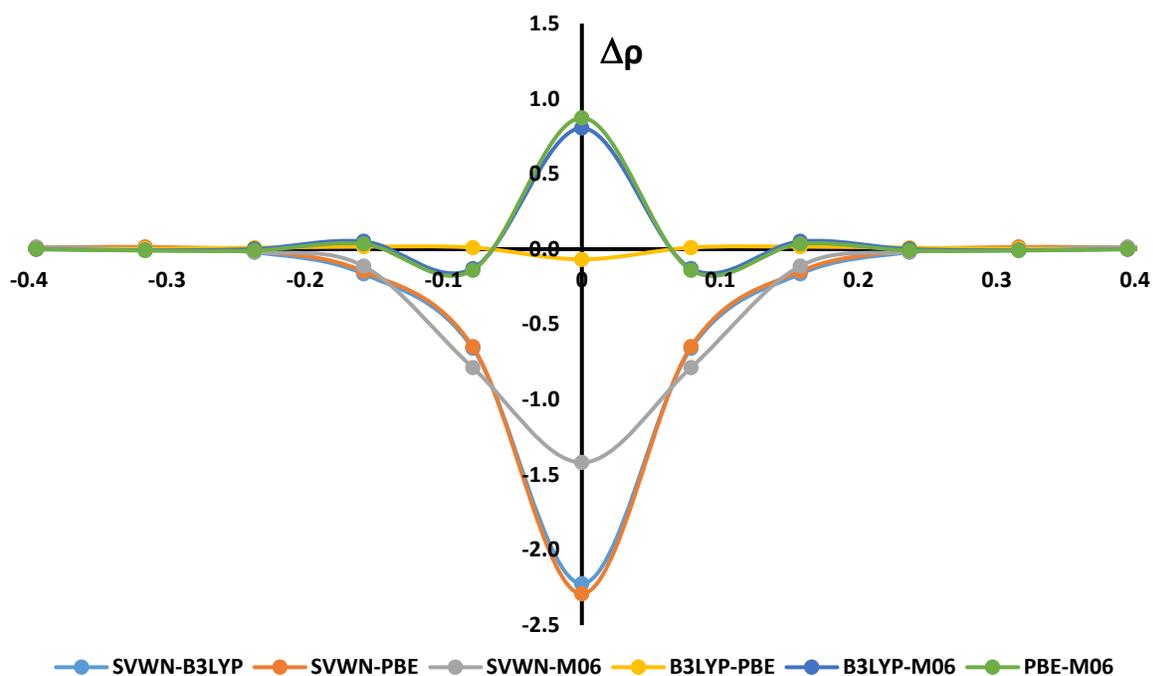

**Figure S4.** The six density difference plots of the four studied density functionals, $C^{4+}$ ion. (aug-cc-pωCV5Z)

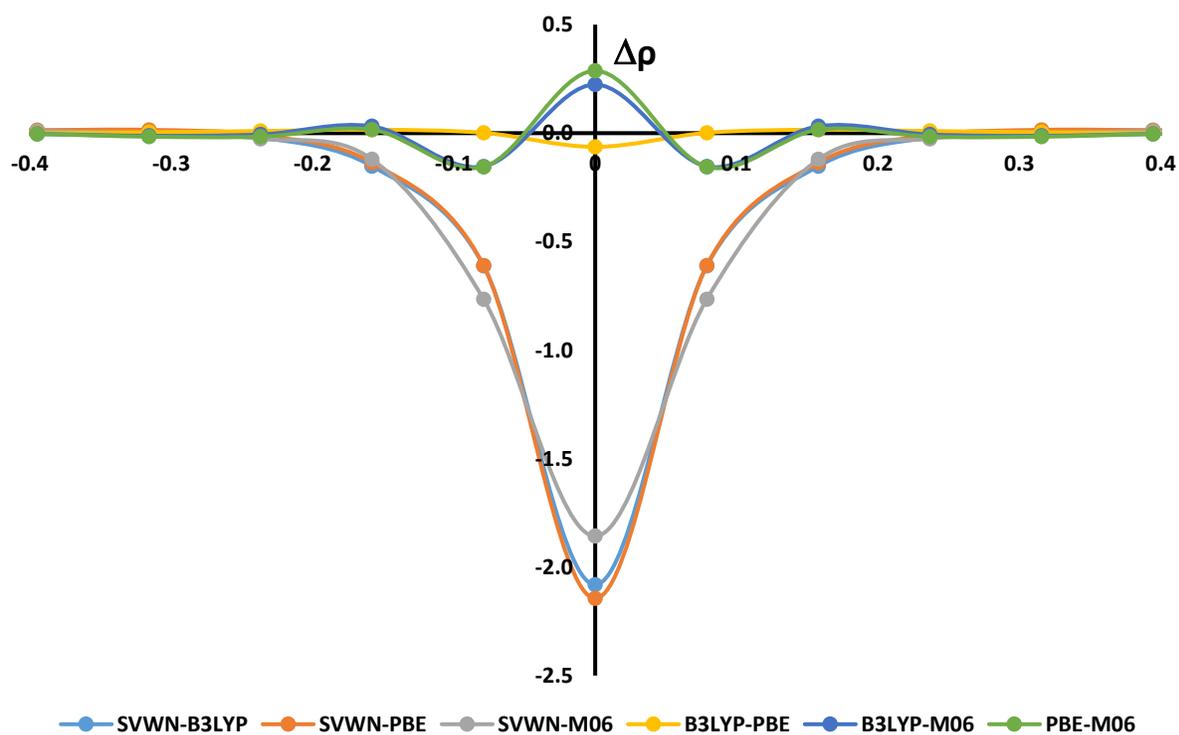

**Figure S5.** The six density difference plots of the four studied density functionals, N$^{3+}$ ion. (aug-cc-pωCV5Z)

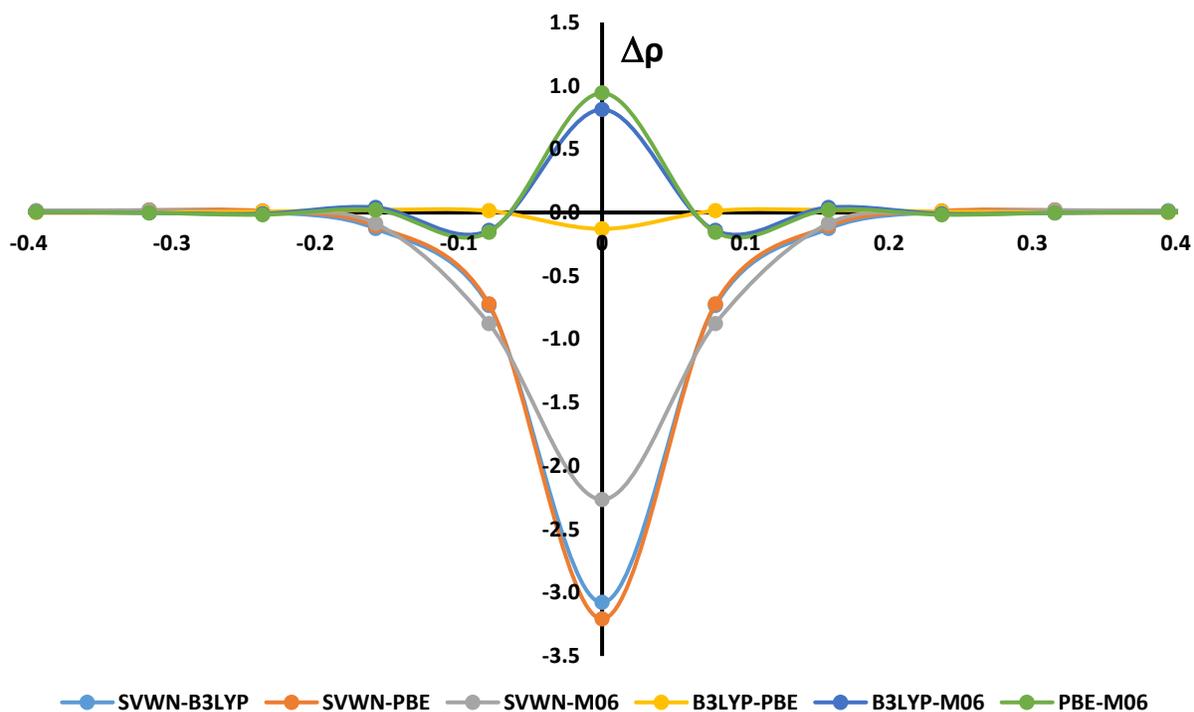

**Figure S6.** The six density difference plots of the four studied density functionals, N$^{5+}$ ion. (aug-cc-pωCV5Z)

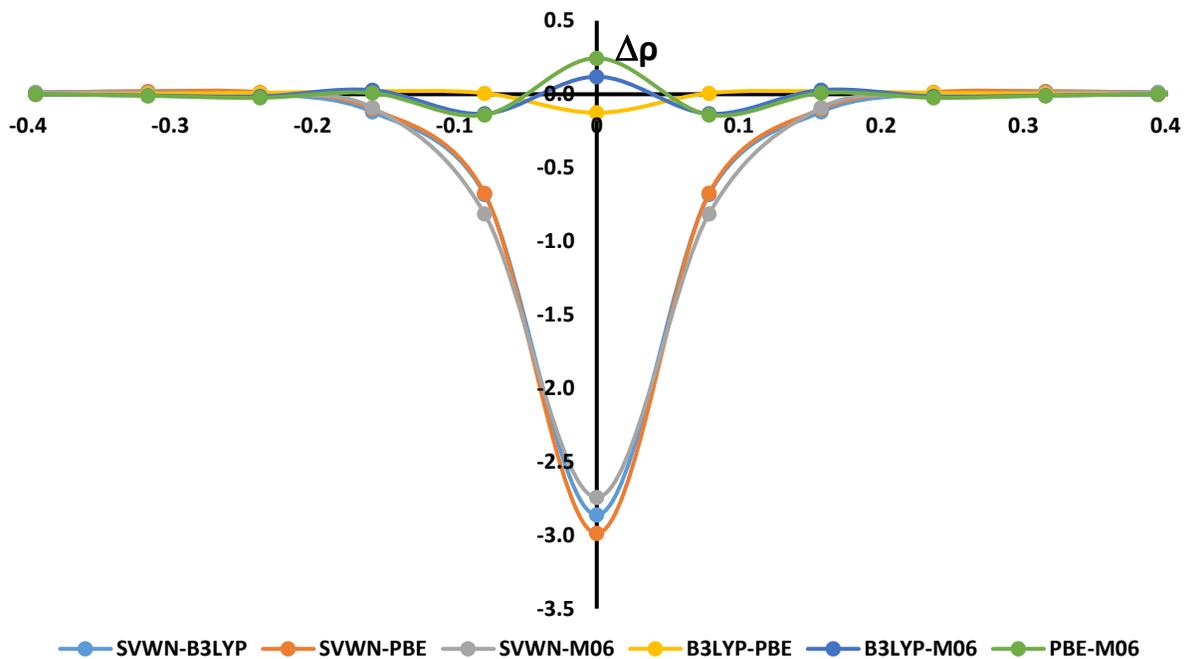

**Figure S7.** The six density difference plots of the four studied density functionals, O$^{4+}$ ion. (aug-cc-pωCV5Z)

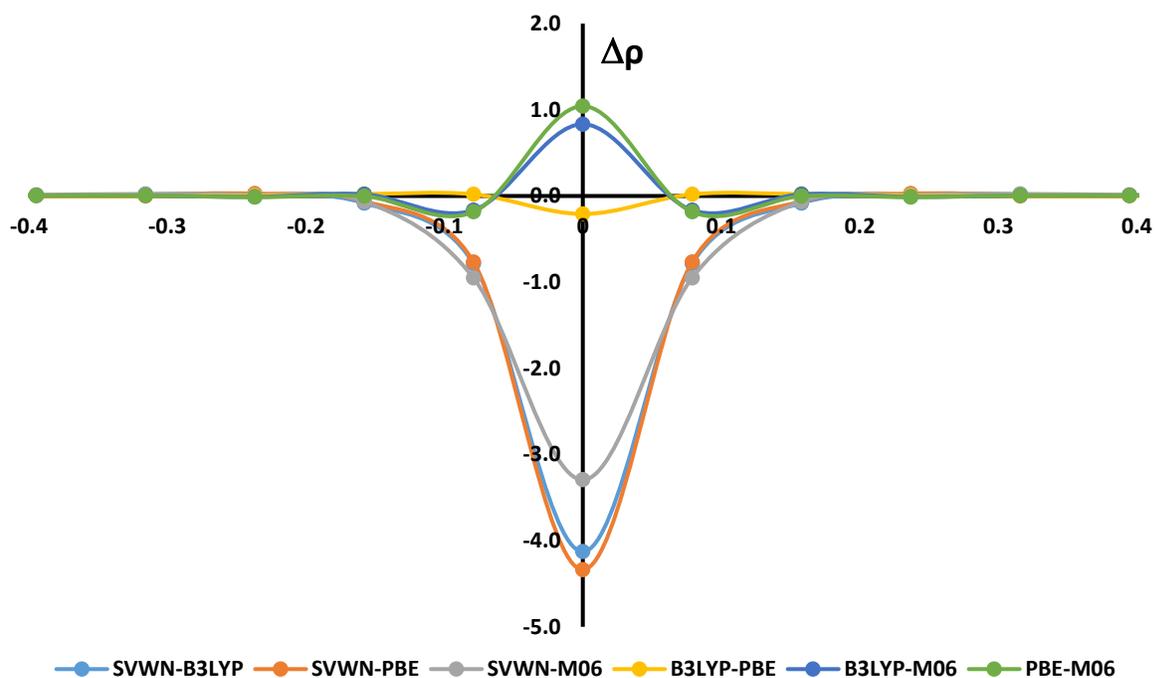

**Figure S8.** The six density difference plots of the four studied density functionals, O$^{6+}$ ion. (aug-cc-pωCV5Z)

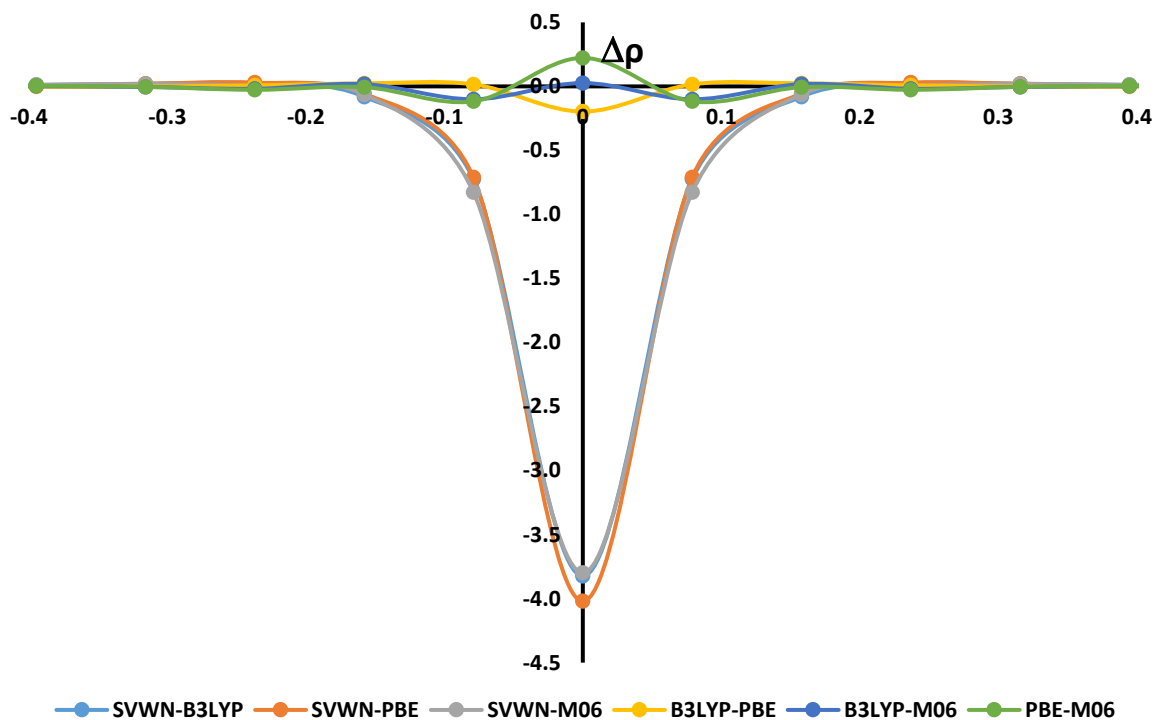

Figure S9. The six density difference plots of the four studied density functionals, $F^{5+}$ ion. (aug-cc-pωCV5Z)

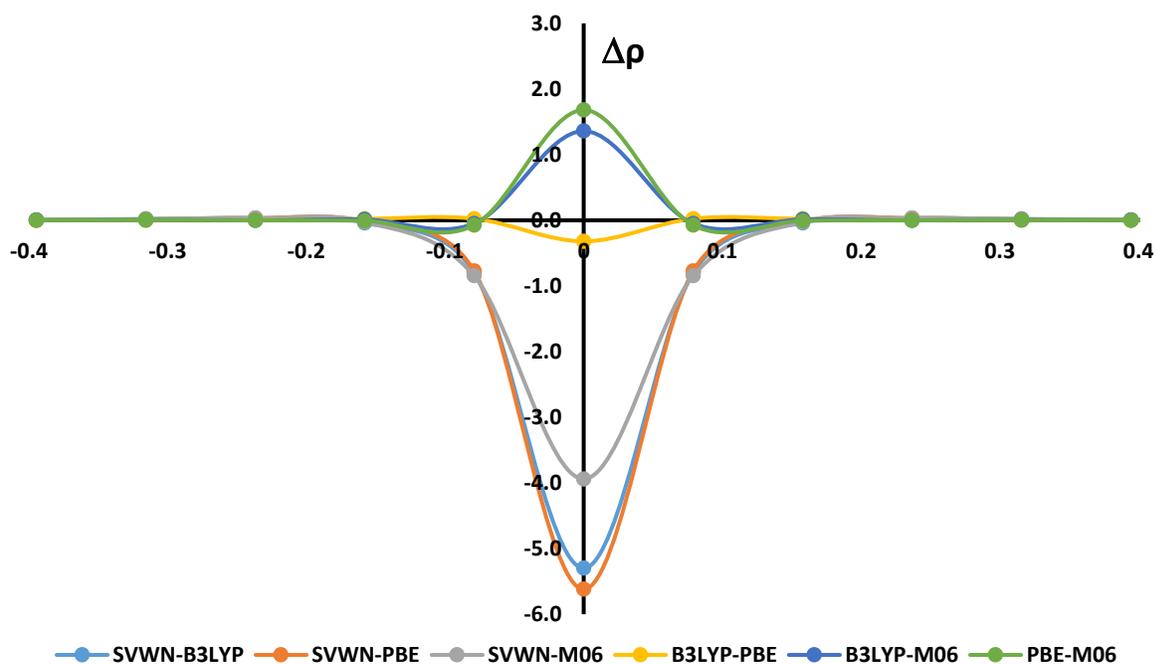

Figure S10. The six density difference plots of the four studied density functionals, $F^{7+}$ ion. (aug-cc-pωCV5Z)

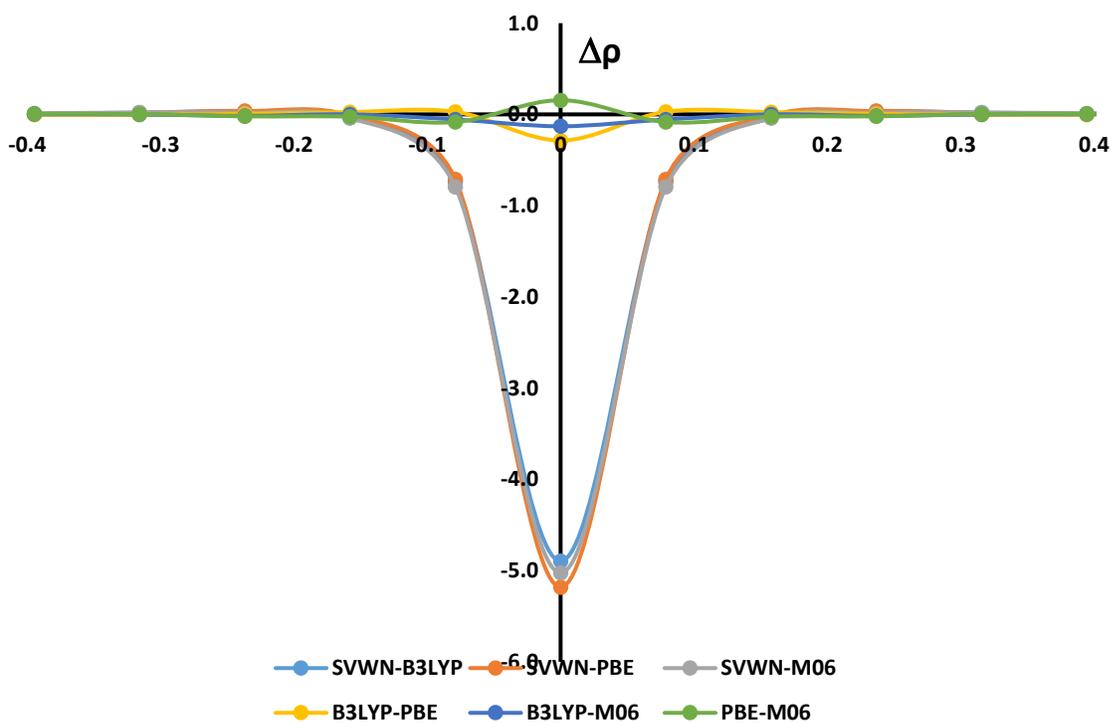

**Figure S11.** The six density difference plots of the four studied density functionals, Ne$^{6+}$ ion. (aug-cc-pωCV5Z)

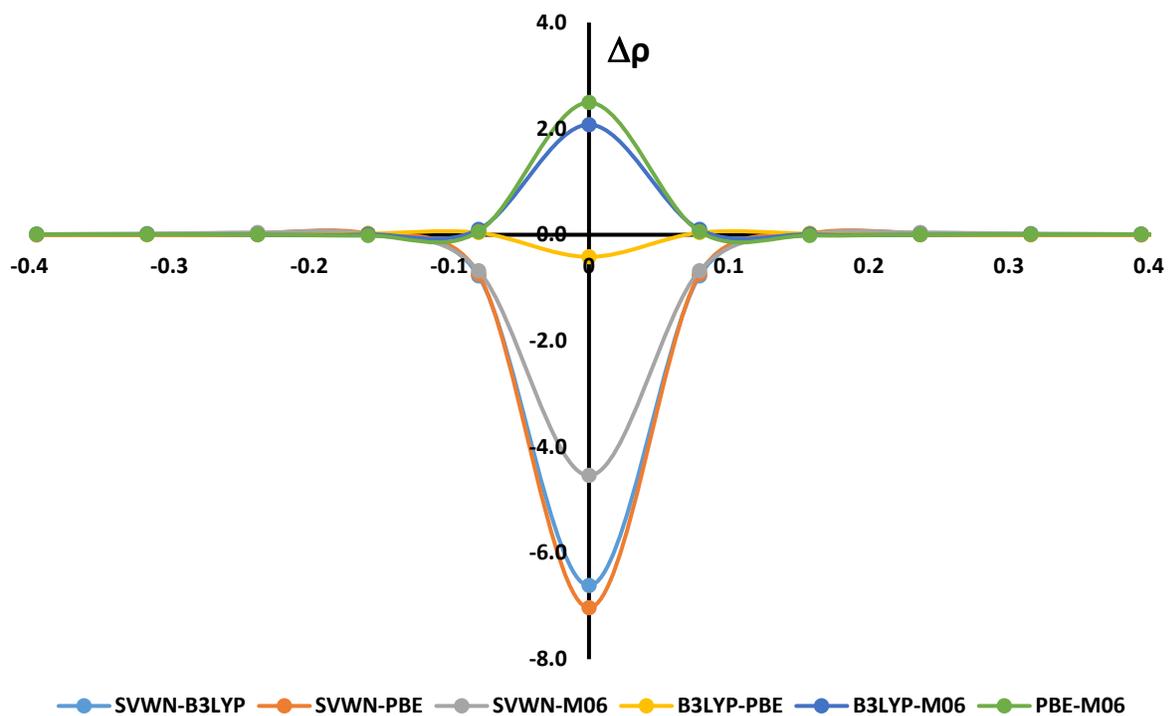

**Figure S12.** The six density difference plots of the four studied density functionals, Ne$^{8+}$ ion. (aug-cc-pωCV5Z)

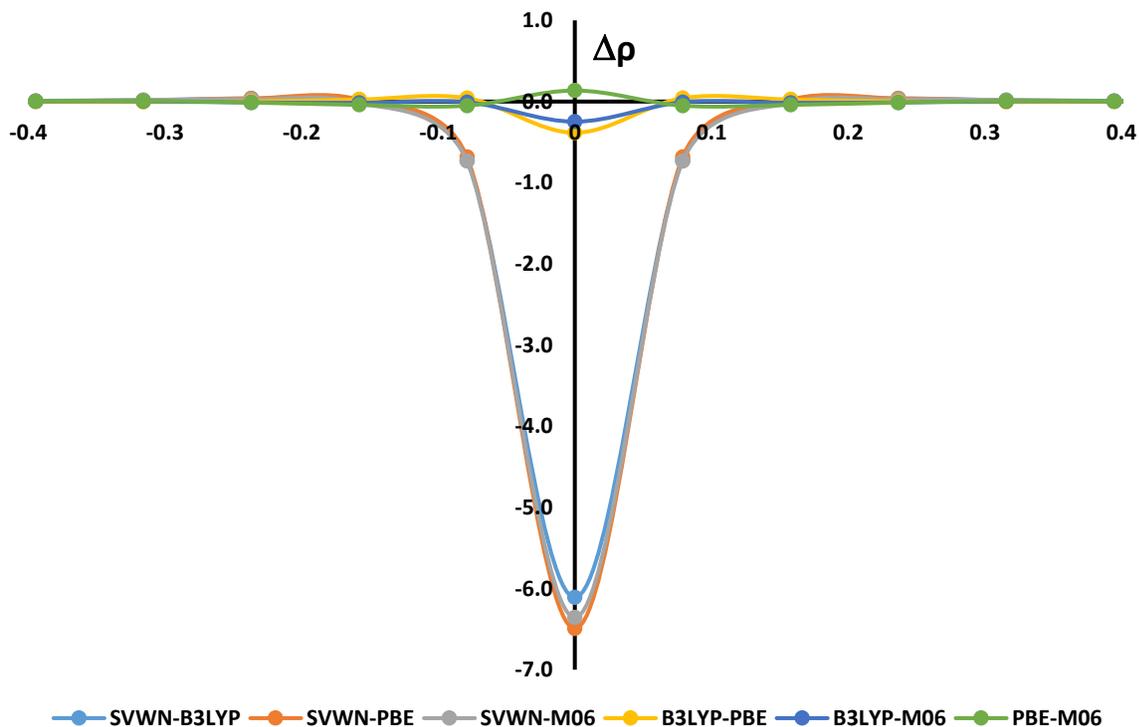

**Figure S13.** Average $\Delta E_D' = E(\rho)-E(\rho')$ in kJ/mol for two test sets: (A) Top: B3LYP, PBE, SVWN, and M06; (B) bottom: PBE, M06-2X, PBE0, and SVWN, using the other three functional's densities as trial densities. The functionals represent very different densities in terms of previous ranking for these systems[1]. The computed energies are the experimentally known double ionization potentials discussed in the main text. Despite the energies being many eVs, these changes in densities affect the computed energies by only < 1.5 kJ/mol, i.e. the density variations manifest below chemical accuracy; such systems are referred to as "practically normal".

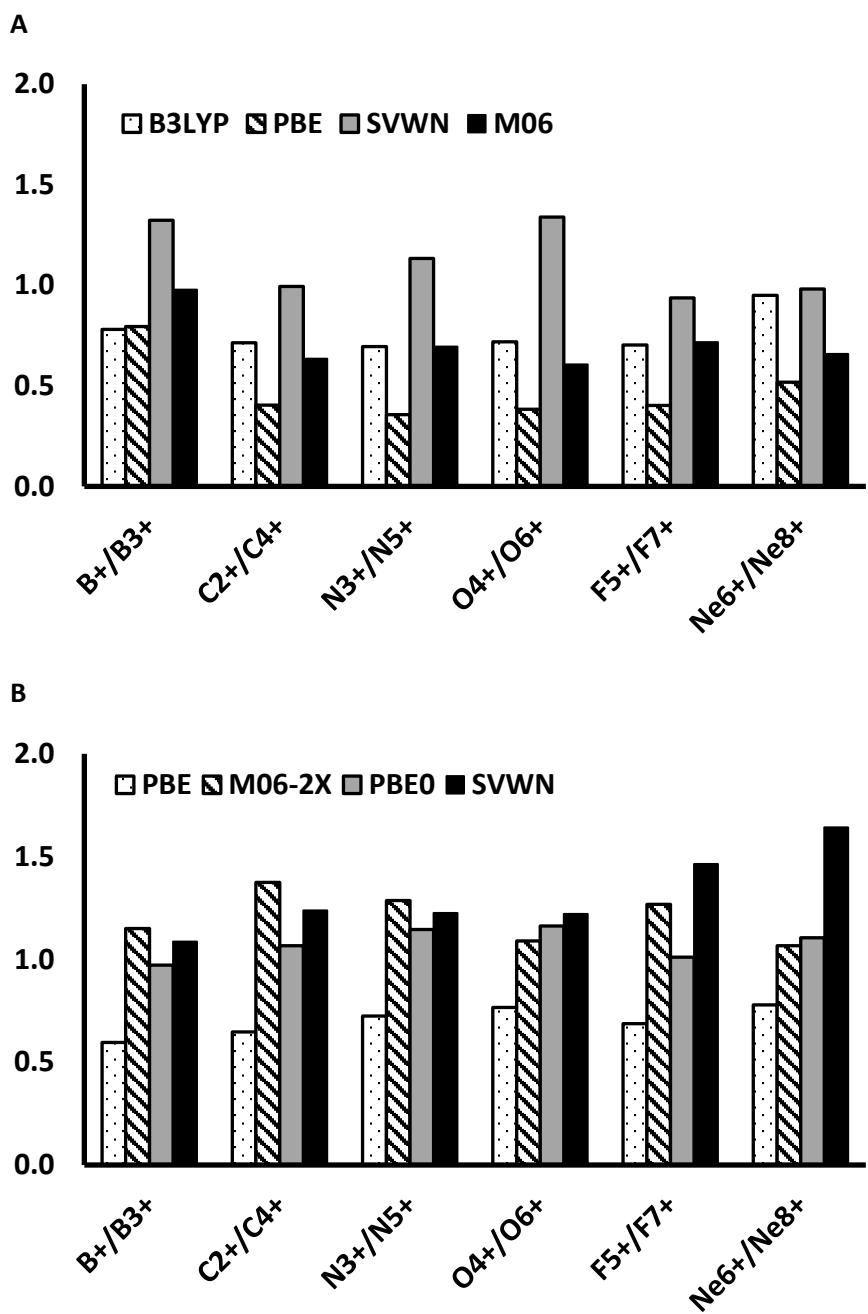

**Figure S14.** Difference between the ionization potentials computed by a given density functional using its own density and other densities: A) M06; B) B3LYP; C) PBE; D) SVWN.

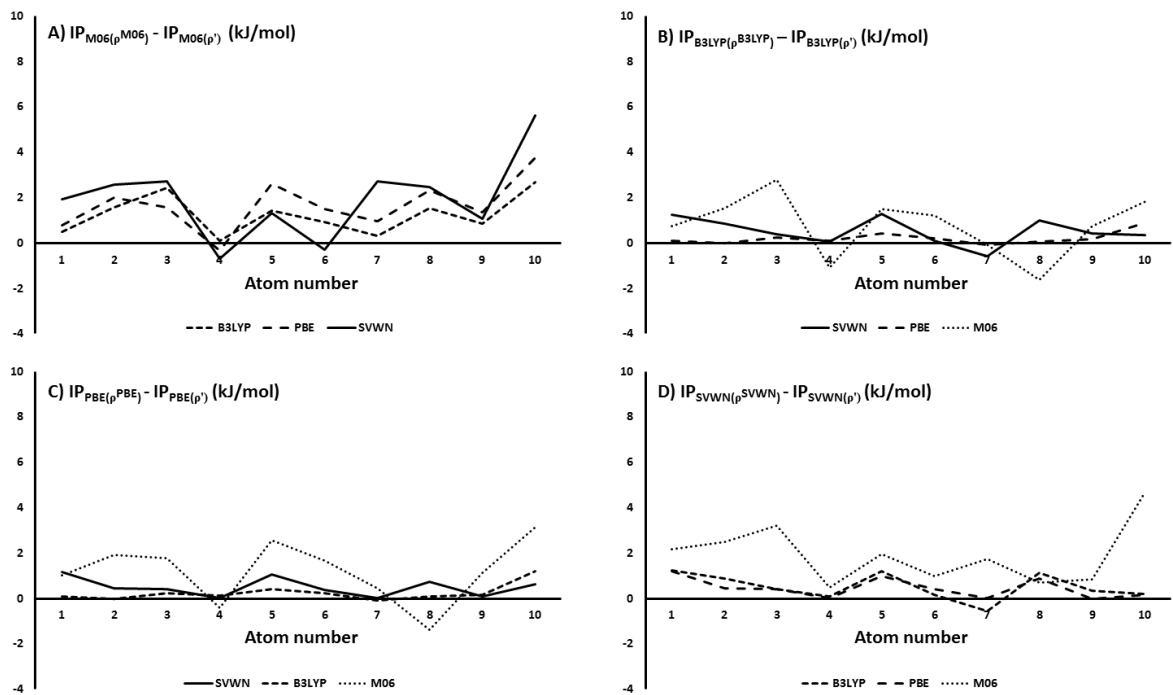

Table S1. Computed B3LYP and PBE energies (a.u.) of ions similar to those studied previously[1,2] (aug-cc-pωCV5Z basis set). The IP values refer to the energy of removing both 2s electrons in kJ/mol. Experimental data from NIST[3].

|  | B3LYP | | | PBE | | | Experiment |
|---|---|---|---|---|---|---|---|
|  | $X^{(n)+}$ | $X^{(n+2)+}$ | IP | $X^{(n)+}$ | $X^{(n+2)+}$ | IP | *(NIST)* |
| $B^+/B^{3+}$ | -24.3298 | -22.0132 | 6082.30 | -24.2934 | -21.9820 | 6068.65 | ***6086.9*** |
| $C^{2+}/C^{4+}$ | -36.5024 | -32.3792 | 10825.34 | -36.4610 | -32.3442 | 10808.71 | ***10843.3*** |
| $N^{3+}/N^{5+}$ | -51.1753 | -44.7448 | 16883.32 | -51.1303 | -44.7063 | 16866.41 | ***16920.3*** |
| $O^{4+}/O^{6+}$ | -68.3483 | -59.1104 | 24253.99 | -68.3006 | -59.0686 | 24238.68 | ***24316.4*** |
| $F^{5+}/F^{7+}$ | -88.0210 | -75.4758 | 32937.47 | -87.9713 | -75.4308 | 32925.06 | ***33032.3*** |
| $Ne^{6+}/Ne^{8+}$ | -110.1936 | -93.8411 | 42933.43 | -110.1422 | -93.7930 | 42924.89 | ***43068.7*** |

Table S2. Computed SVWN and M06 energies (a.u.) of ions similar to those studied previously[1,2] (aug-cc-pωCV5Z basis set). The IP values refer to the energy of removing both 2s electrons in kJ/mol.

|  | SVWN | | | M06 | | |
|---|---|---|---|---|---|---|
|  | $X^{(n)+}$ | $X^{(n+2)+}$ | IP | $X^{(n)+}$ | $X^{(n+2)+}$ | IP |
| $B^+/B^{3+}$ | -24.0382 | -21.7432 | 6025.72 | -24.3340 | -22.0492 | 5998.70 |
| $C^{2+}/C^{4+}$ | -36.1300 | -32.0395 | 10739.70 | -36.5087 | -32.4265 | 10717.78 |
| $N^{3+}/N^{5+}$ | -50.7211 | -44.3342 | 16768.75 | -51.1843 | -44.8034 | 16753.11 |
| $O^{4+}/O^{6+}$ | -67.8112 | -58.6282 | 24109.87 | -68.3598 | -59.1802 | 24101.01 |
| $F^{5+}/F^{7+}$ | -87.4003 | -74.9213 | 32763.70 | -88.0353 | -75.5567 | 32762.41 |
| $Ne^{6+}/Ne^{8+}$ | -109.4886 | -93.2137 | 42729.73 | -110.2103 | -93.9331 | 42735.76 |

**Table S3.** Computed PBE, B3LYP, and M06 energies (a.u.) using the optimized SVWN densities of ions similar to those studied previously[1,2] (aug-cc-pωCV5Z basis set). The IP values refer to the energy of removing both 2s electrons in kJ/mol.

|  | PBE(SVWN) | | | B3LYP(SVWN) | | | M06(SVWN) | | |
| --- | --- | --- | --- | --- | --- | --- | --- | --- | --- |
|  | $X^{(n)+}$ | $X^{(n+2)+}$ | IP | $X^{(n)+}$ | $X^{(n+2)+}$ | IP | $X^{(n)+}$ | $X^{(n+2)+}$ | IP |
| $B^+/B^{3+}$ | -24.2920 | -21.9808 | 6067.93 | -24.3283 | -22.0119 | 6081.62 | -24.3312 | -22.0469 | 5997.38 |
| $C^{2+}/C^{4+}$ | -36.4595 | -32.3430 | 10807.96 | -36.5008 | -32.3779 | 10824.60 | -36.5061 | -32.4243 | 10716.77 |
| $N^{3+}/N^{5+}$ | -51.1288 | -44.7050 | 16865.60 | -51.1737 | -44.7435 | 16882.52 | -51.1818 | -44.8012 | 16752.17 |
| $O^{4+}/O^{6+}$ | -68.2990 | -59.0673 | 24237.79 | -68.3466 | -59.1091 | 24253.13 | -68.3574 | -59.1780 | 24100.35 |
| $F^{5+}/F^{7+}$ | -87.9696 | -75.4295 | 32924.12 | -88.0193 | -75.4744 | 32936.57 | -88.0327 | -75.5546 | 32761.41 |
| $Ne^{6+}/Ne^{8+}$ | -110.1405 | -93.7917 | 42923.91 | -110.1919 | -93.8397 | 42932.50 | -110.2078 | -93.9309 | 42734.88 |

**Table S4.** Computed SVWN, B3LYP, and M06 energies (a.u.) using the optimized PBE densities of ions similar to those studied previously[1,2] (aug-cc-pωCV5Z basis set). The IP values refer to the energy of removing both 2s electrons in kJ/mol.

|  | SVWN(PBE) | | | B3LYP(PBE) | | | M06(PBE) | | |
| --- | --- | --- | --- | --- | --- | --- | --- | --- | --- |
|  | $X^{(n)+}$ | $X^{(n+2)+}$ | IP | $X^{(n)+}$ | $X^{(n+2)+}$ | IP | $X^{(n)+}$ | $X^{(n+2)+}$ | IP |
| $B^+/B^{3+}$ | -24.0368 | -21.7420 | 6025.00 | -24.3296 | -22.0131 | 6081.93 | -24.3328 | -22.0482 | 5998.04 |
| $C^{2+}/C^{4+}$ | -36.1285 | -32.0383 | 10738.95 | -36.5022 | -32.3791 | 10825.06 | -36.5079 | -32.4257 | 10718.01 |
| $N^{3+}/N^{5+}$ | -50.7195 | -44.3330 | 16767.94 | -51.1751 | -44.7447 | 16883.10 | -51.1837 | -44.8026 | 16753.56 |
| $O^{4+}/O^{6+}$ | -67.8096 | -58.6269 | 24108.96 | -68.3481 | -59.1103 | 24253.80 | -68.3592 | -59.1794 | 24101.56 |
| $F^{5+}/F^{7+}$ | -87.3987 | -74.9200 | 32762.74 | -88.0209 | -75.4757 | 32937.34 | -88.0346 | -75.5560 | 32762.65 |
| $Ne^{6+}/Ne^{8+}$ | -109.4869 | -93.2124 | 42728.76 | -110.1934 | -93.8410 | 42933.31 | -110.2096 | -93.9323 | 42736.07 |

**Table S5.** Computed SVWN, PBE, and M06 energies (a.u.) using the optimized B3LYP densities of ions similar to those studied previously[1,2] (aug-cc-pωCV5Z basis set). The IP values refer to the energy of removing both 2s electrons in kJ/mol.

|  | SVWN(B3LYP) | | | PBE (B3LYP) | | | M06(B3LYP) | | |
|---|---|---|---|---|---|---|---|---|---|
|  | $X^{(n)+}$ | $X^{(n+2)+}$ | IP | $X^{(n)+}$ | $X^{(n+2)+}$ | IP | $X^{(n)+}$ | $X^{(n+2)+}$ | IP |
| $B^+/B^{3+}$ | -24.0367 | -21.7419 | 6025.04 | -24.2932 | -21.9819 | 6068.30 | -24.3328 | -22.0484 | 5997.75 |
| $C^{2+}/C^{4+}$ | -36.1285 | -32.0382 | 10738.96 | -36.4608 | -32.3441 | 10808.44 | -36.5078 | -32.4259 | 10717.12 |
| $N^{3+}/N^{5+}$ | -50.7195 | -44.3329 | 16767.96 | -51.1302 | -44.7062 | 16866.19 | -51.1835 | -44.8028 | 16752.43 |
| $O^{4+}/O^{6+}$ | -67.8095 | -58.6269 | 24109.00 | -68.3004 | -59.0685 | 24238.51 | -68.3590 | -59.1797 | 24100.40 |
| $F^{5+}/F^{7+}$ | -87.3986 | -74.9199 | 32762.80 | -87.9711 | -75.4307 | 32924.92 | -88.0344 | -75.5563 | 32761.50 |
| $Ne^{6+}/Ne^{8+}$ | -109.4868 | -93.2123 | 42728.80 | -110.1421 | -93.7929 | 42924.78 | -110.2095 | -93.9326 | 42734.98 |

**Table S6.** Computed B3LYP, PBE, and SVWN energies (a.u.) using the optimized M06 densities of ions similar to those studied previously[1,2] (aug-cc-pωCV5Z basis set). The IP values refer to the energy of removing both 2s electrons in kJ/mol.

|  | B3LYP(M06) | | | PBE(M06) | | | SVWN(M06) | | |
|---|---|---|---|---|---|---|---|---|---|
|  | $X^{(n)+}$ | $X^{(n+2)+}$ | IP | $X^{(n)+}$ | $X^{(n+2)+}$ | IP | $X^{(n)+}$ | $X^{(n+2)+}$ | IP |
| $B^+/B^{3+}$ | -24.3285 | -22.0124 | 6081.01 | -24.2919 | -21.9810 | 6067.34 | -24.0350 | -21.7409 | 6023.15 |
| $C^{2+}/C^{4+}$ | -36.5013 | -32.3786 | 10824.23 | -36.4601 | -32.3433 | 10808.53 | -36.1273 | -32.0373 | 10738.20 |
| $N^{3+}/N^{5+}$ | -51.1744 | -44.7443 | 16882.25 | -51.1296 | -44.7055 | 16866.45 | -50.7183 | -44.3321 | 16766.96 |
| $O^{4+}/O^{6+}$ | -68.3473 | -59.1099 | 24252.88 | -68.2998 | -59.0678 | 24238.59 | -67.8082 | -58.6261 | 24107.62 |
| $F^{5+}/F^{7+}$ | -88.0202 | -75.4753 | 32936.38 | -87.9705 | -75.4300 | 32924.93 | -87.3978 | -74.9191 | 32762.76 |
| $Ne^{6+}/Ne^{8+}$ | -110.1925 | -93.8407 | 42931.63 | -110.1413 | -93.7923 | 42924.41 | -109.4860 | -93.2115 | 42728.68 |

**Table S7. Experimental ionization potentials and ground states of atoms and ions from NIST.**

| ATOM | Exp. IP/eV | STATE X | STATE X$^+$ |
|---|---|---|---|
| H | 13.60 | $^2S_{1/2}$ | |
| He | 24.59 | $^1S_0$ | $^2S_{1/2}$ |
| Li | 5.39 | $^2S_{1/2}$ | $^1S_0$ |
| Be | 9.32 | $^1S_0$ | $^2S_{1/2}$ |
| B | 8.30 | $^2P°_{1/2}$ | $^1S_0$ |
| C | 11.26 | $^3P_0$ | $^2P°_{1/2}$ |
| N | 14.53 | $^4S°_{3/2}$ | $^3P_0$ |
| O | 13.62 | $^3P_2$ | $^4S°_{3/2}$ |
| F | 17.42 | $^2P°_{3/2}$ | $^3P_2$ |
| Ne | 21.56 | $^1S_0$ | $^2P°_{3/2}$ |

**Table S8. Electronic energies for atoms and ions (in a.u.) and ionization potentials (IP) for atoms (in kJ/mol), computed using CCSD(T) and CCSD, aug-cc-pV5Z basis set.**

| ATOM | CCSD(T) | | | CCSD | | |
|---|---|---|---|---|---|---|
| | X | X+ | IP | X | X+ | IP |
| H | -0.5000 | 0.0000 | **1312.74** | -0.5000 | 0.0000 | **1312.74** |
| He | -2.9032 | -1.9999 | **2371.50** | -2.9032 | -1.9999 | **2371.50** |
| Li | -7.4600 | -7.2622 | **519.15** | -7.4599 | -7.2622 | **519.07** |
| Be | -14.6463 | -14.3042 | **898.23** | -14.6458 | -14.3042 | **896.82** |
| B | -24.6298 | -24.3255 | **799.01** | -24.6279 | -24.3249 | **795.64** |
| C | -37.8195 | -37.4060 | **1085.58** | -37.8167 | -37.4043 | **1082.81** |
| N | -54.5627 | -54.0284 | **1402.58** | -54.5594 | -54.0262 | **1399.85** |
| O | -75.0372 | -74.5384 | **1309.40** | -75.0329 | -74.5362 | **1304.06** |
| F | -99.7000 | -99.0607 | **1678.73** | -99.6946 | -99.0577 | **1672.20** |
| Ne | -128.9004 | -128.1074 | **2082.13** | -128.8939 | -128.1037 | **2074.66** |

Table S9. Electronic energies for atoms and ions (in a.u.) and ionization potentials (IP) for atoms kJ/mol), computed using B3LYP and PBE functionals, aug-cc-pV5Z basis set.

| ATOM | B3LYP | | | PBE | | |
|---|---|---|---|---|---|---|
| | X | X+ | IP | X | X+ | IP |
| H | -0.4991 | 0.0000 | **1310.26** | -0.5000 | 0.0000 | **1312.71** |
| He | -2.9081 | -1.9949 | **2397.62** | -2.8929 | -1.9937 | **2360.75** |
| Li | -7.4825 | -7.2786 | **535.44** | -7.4620 | -7.2567 | **539.00** |
| Be | -14.6593 | -14.3276 | **871.05** | -14.6298 | -14.2992 | **868.05** |
| B | -24.6479 | -24.3297 | **835.39** | -24.6120 | -24.2932 | **837.04** |
| C | -37.8402 | -37.4192 | **1105.43** | -37.7985 | -37.3743 | **1113.55** |
| N | -54.5818 | -54.0464 | **1405.79** | -54.5354 | -53.9943 | **1420.76** |
| O | -75.0712 | -74.5556 | **1353.91** | -75.0146 | -74.4978 | **1356.73** |
| F | -99.7421 | -99.0947 | **1699.95** | -99.6756 | -99.0269 | **1703.26** |
| Ne | -128.9422 | -128.1483 | **2084.27** | -128.8658 | -128.0700 | **2089.26** |

Table S10. Electronic energies for atoms and ions (in a.u.) and ionization potentials (IP) for atoms kJ/mol), computed using SVWN and M06 functionals, aug-cc-pV5Z basis set.

| ATOM | SVWN | | | M06 | | |
|---|---|---|---|---|---|---|
| | X | X+ | IP | X | X+ | IP |
| H | -0.4787 | 0.0000 | **1256.74** | -0.5002 | 0.0000 | **1313.25** |
| He | -2.8348 | -1.9417 | **2344.93** | -2.9102 | -1.9996 | **2390.90** |
| Li | -7.3439 | -7.1427 | **528.11** | -7.4859 | -7.2908 | **512.06** |
| Be | -14.4471 | -14.1154 | **870.93** | -14.6609 | -14.3325 | **862.23** |
| B | -24.3560 | -24.0381 | **834.49** | -24.6425 | -24.3325 | **813.85** |
| C | -37.4701 | -37.0404 | **1128.22** | -37.8305 | -37.4145 | **1092.34** |
| N | -54.1365 | -53.5852 | **1447.62** | -54.5789 | -54.0373 | **1421.95** |
| O | -74.5310 | -74.0163 | **1351.17** | -75.0651 | -74.5576 | **1332.58** |
| F | -99.1144 | -98.4545 | **1732.62** | -99.7399 | -99.0950 | **1693.17** |
| Ne | -128.2329 | -127.4176 | **2140.53** | -128.9512 | -128.1549 | **2090.93** |

Table S11. Electronic energies for atoms and ions (in a.u.) and ionization potentials (IP) for atoms kJ/mol), computed using SWN densities but other functionals, aug-cc-pV5Z basis set.

| ATOM | PBE(SVWN) | | | B3LYP(SVWN) | | | M06(SVWN) | | |
|---|---|---|---|---|---|---|---|---|---|
| | X | X+ | IP | X | X+ | IP | X | X+ | IP |
| H | -0.4996 | 0.0000 | **1311.6** | -0.4986 | 0.0000 | **1309.1** | -0.4995 | 0.0000 | **1311.4** |
| He | -2.8922 | -1.9932 | **2360.3** | -2.9072 | -1.9943 | **2396.7** | -2.9084 | -1.9988 | **2388.3** |
| Li | -7.4609 | -7.2558 | **538.6** | -7.4813 | -7.2776 | **535.0** | -7.4830 | -7.2894 | **508.3** |
| Be | -14.6286 | -14.2980 | **868.1** | -14.6579 | -14.3262 | **871.0** | -14.6586 | -14.3313 | **859.2** |
| B | -24.6102 | -24.2920 | **835.6** | -24.6459 | -24.3282 | **834.1** | -24.6385 | -24.3308 | **807.8** |
| C | -37.7964 | -37.3725 | **1112.9** | -37.8381 | -37.4171 | **1105.3** | -37.8272 | -37.4123 | **1089.3** |
| N | -54.5333 | -53.9922 | **1420.6** | -54.5798 | -54.0441 | **1406.3** | -54.5727 | -54.0344 | **1413.3** |
| O | -75.0121 | -74.4956 | **1356.1** | -75.0687 | -74.5534 | **1353.0** | -75.0615 | -74.5552 | **1329.3** |
| F | -99.6732 | -99.0246 | **1703.0** | -99.7395 | -99.0923 | **1699.4** | -99.7341 | -99.0906 | **1689.5** |
| Ne | -128.8639 | -128.0680 | **2089.8** | -128.9400 | -128.1460 | **2084.7** | -128.9479 | -128.1512 | **2091.5** |

Table S12. Electronic energies for atoms and ions (in a.u.) and ionization potentials (IP) for atoms (in kJ/mol), computed using PBE densities but other functionals, aug-cc-pV5Z basis set.

| ATOM | SVWN(PBE) | | | B3LYP(PBE) | | | M06(PBE) | | |
|---|---|---|---|---|---|---|---|---|---|
| | X | X+ | IP | X | X+ | IP | X | X+ | IP |
| H | -0.4782 | 0.0000 | **1255.5** | -0.4990 | 0.0000 | **1310.2** | -0.4999 | 0.0000 | **1312.5** |
| He | -2.8341 | -1.9411 | **2344.5** | -2.9080 | -1.9948 | **2397.6** | -2.9093 | -1.9994 | **2388.9** |
| Li | -7.3428 | -7.1418 | **527.7** | -7.4824 | -7.2786 | **535.2** | -7.4844 | -7.2906 | **509.1** |
| Be | -14.4459 | -14.1142 | **870.9** | -14.6592 | -14.3274 | **870.9** | -14.6604 | -14.3327 | **860.2** |
| B | -24.3541 | -24.0367 | **833.5** | -24.6475 | -24.3295 | **834.9** | -24.6400 | -24.3327 | **806.6** |
| C | -37.4681 | -37.0385 | **1127.8** | -37.8398 | -37.4189 | **1105.2** | -37.8286 | -37.4140 | **1088.4** |
| N | -54.1344 | -53.5831 | **1447.6** | -54.5815 | -54.0460 | **1405.8** | -54.5750 | -54.0361 | **1414.8** |
| O | -74.5285 | -74.0142 | **1350.3** | -75.0709 | -74.5552 | **1353.9** | -75.0637 | -74.5568 | **1330.9** |
| F | -99.1121 | -98.4522 | **1732.6** | -99.7420 | -99.0946 | **1699.7** | -99.7367 | -99.0933 | **1689.2** |
| Ne | -128.2308 | -127.4153 | **2141.0** | -128.9423 | -128.1484 | **2084.2** | -128.9514 | -128.1540 | **2093.8** |

**Table S13.** Electronic energies for atoms and ions (in a.u.) and ionization potentials (IP) for atoms (in kJ/mol), computed using B3LYP densities but other functionals, aug-cc-pV5Z basis set.

| ATOM | SVWN (B3LYP) | | | PBE (B3LYP) | | | M06 (B3LYP) | | |
|---|---|---|---|---|---|---|---|---|---|
| | X | X+ | IP | X | X+ | IP | X | X+ | IP |
| H | -0.4782 | 0.0000 | **1255.5** | -0.5000 | 0.0000 | **1312.6** | -0.4991 | 0.0000 | **1310.3** |
| He | -2.8339 | -1.9411 | **2344.1** | -2.8928 | -1.9937 | **2360.7** | -2.9095 | -1.9995 | **2389.3** |
| Li | -7.3427 | -7.1417 | **527.7** | -7.4619 | -7.2567 | **538.7** | -7.4843 | -7.2907 | **508.3** |
| Be | -14.4457 | -14.1140 | **870.8** | -14.6297 | -14.2991 | **868.0** | -14.6604 | -14.3325 | **860.8** |
| B | -24.3540 | -24.0366 | **833.3** | -24.6117 | -24.2932 | **836.2** | -24.6404 | -24.3327 | **807.9** |
| C | -37.4679 | -37.0383 | **1128.0** | -37.7980 | -37.3741 | **1113.1** | -37.8290 | -37.4142 | **1089.0** |
| N | -54.1345 | -53.5829 | **1448.2** | -54.5350 | -53.9939 | **1420.6** | -54.5754 | -54.0364 | **1415.2** |
| O | -74.5284 | -74.0143 | **1350.0** | -75.0143 | -74.4975 | **1356.8** | -75.0644 | -74.5573 | **1331.3** |
| F | -99.1119 | -98.4521 | **1732.3** | -99.6754 | -99.0268 | **1702.9** | -99.7373 | -99.0937 | **1689.7** |
| Ne | -128.2306 | -127.4152 | **2140.9** | -128.8660 | -128.0703 | **2089.3** | -128.9522 | -128.1543 | **2094.9** |

**Table S14.** Electronic energies for atoms and ions (in a.u.) and ionization potentials (IP) for atoms (in kJ/mol), computed using M06 densities but other functionals, aug-cc-pV5Z basis set.

| ATOM | B3LYP(M06) | | | PBE(M06) | | | SVWN(M06) | | |
|---|---|---|---|---|---|---|---|---|---|
| | X | X+ | IP | X | X+ | IP | X | X+ | IP |
| H | -0.4988 | 0.0000 | **1309.5** | -0.4996 | 0.0000 | **1311.7** | -0.4778 | 0.0000 | **1254.6** |
| He | -2.9074 | -1.9947 | **2396.1** | -2.8920 | -1.9935 | **2358.8** | -2.8330 | -1.9408 | **2342.4** |
| Li | -7.4813 | -7.2784 | **532.7** | -7.4610 | -7.2563 | **537.3** | -7.3412 | -7.1412 | **525.0** |
| Be | -14.6589 | -14.3267 | **872.1** | -14.6294 | -14.2986 | **868.5** | -14.4449 | -14.1133 | **870.4** |
| B | -24.6469 | -24.3293 | **833.8** | -24.6107 | -24.2930 | **834.0** | -24.3529 | -24.0359 | **832.5** |
| C | -37.8389 | -37.4184 | **1104.2** | -37.7968 | -37.3733 | **1111.7** | -37.4669 | -37.0375 | **1127.2** |
| N | -54.5808 | -54.0454 | **1405.8** | -54.5339 | -53.9930 | **1420.1** | -54.1328 | -53.5821 | **1445.9** |
| O | -75.0698 | -74.5550 | **1351.6** | -75.0125 | -74.4952 | **1358.2** | -74.5268 | -74.0124 | **1350.5** |
| F | -99.7409 | -99.0937 | **1699.1** | -99.6737 | -99.0255 | **1701.9** | -99.1101 | -98.4505 | **1731.8** |
| Ne | -128.9409 | -128.1474 | **2083.3** | -128.8639 | -128.0689 | **2087.3** | -128.2273 | -127.4136 | **2136.4** |

Table S15. Electronic energies for atoms and ions (in a.u.) and ionization potentials (IP) for atoms (in kJ/mol), computed using B3LYP and PBE functionals, def-TZVP basis set.

| ATOM | B3LYP | | | PBE | | |
|---|---|---|---|---|---|---|
| | X | X+ | IP | X | X+ | IP |
| H | -0.4988 | 0.0000 | **1309.51** | -0.4996 | 0.0000 | **1311.74** |
| He | -2.9059 | -1.9933 | **2396.12** | -2.8906 | -1.9921 | **2358.97** |
| Li | -7.4824 | -7.2785 | **535.32** | -7.4619 | -7.2566 | **538.98** |
| Be | -14.6581 | -14.3264 | **870.86** | -14.6283 | -14.2977 | **867.93** |
| B | -24.6463 | -24.3283 | **834.93** | -24.6101 | -24.2915 | **836.54** |
| C | -37.8382 | -37.4173 | **1104.84** | -37.7960 | -37.3722 | **1112.73** |
| N | -54.5790 | -54.0435 | **1405.88** | -54.5322 | -53.9912 | **1420.35** |
| O | -75.0669 | -74.5523 | **1351.01** | -75.0097 | -74.4943 | **1353.03** |
| F | -99.7363 | -99.0898 | **1697.35** | -99.6691 | -99.0218 | **1699.64** |
| Ne | -0.4988 | 0.0000 | **1309.51** | -0.4996 | 0.0000 | **1311.74** |

Table S16. Electronic energies for atoms and ions (in a.u.) and ionization potentials (IP) for atoms (in kJ/mol), computed using SVWN and M06 functionals, def-TZVP basis set.

| ATOM | SVWN | | | M06 | | |
|---|---|---|---|---|---|---|
| | X | X+ | IP | X | X+ | IP |
| H | -0.4783 | 0.0000 | **1255.90** | -0.4999 | 0.0000 | **1312.41** |
| He | -2.8326 | -1.9402 | **2342.98** | -2.9082 | -1.9990 | **2386.98** |
| Li | -7.3438 | -7.1427 | **527.88** | -7.4861 | -7.2911 | **512.08** |
| Be | -14.4457 | -14.1142 | **870.31** | -14.6596 | -14.3316 | **861.19** |
| B | -24.3539 | -24.0365 | **833.54** | -24.6407 | -24.3314 | **812.16** |
| C | -37.4674 | -37.0381 | **1127.15** | -37.8279 | -37.4123 | **1091.05** |
| N | -54.1329 | -53.5817 | **1446.99** | -54.5765 | -54.0342 | **1423.70** |
| O | -74.5256 | -74.0124 | **1347.36** | -75.0612 | -74.5552 | **1328.50** |
| F | -99.1074 | -98.4488 | **1729.30** | -99.7345 | -99.0911 | **1689.47** |
| Ne | -128.2241 | -127.4090 | **2140.11** | -128.9457 | -128.1492 | **2091.12** |

**Table S17.** Errors in computed ionization potentials (in kJ/mol) with two basis sets as specified.

|     | aug-cc-pV5Z | | | | def-TZVP | | | |
|-----|-------|-------|-------|-------|-------|-------|-------|-------|
|     | B3LYP | PBE   | SVWN  | M06   | B3LYP | PBE   | SVWN  | M06   |
| H   | 1.81  | -0.64 | 55.34 | -1.18 | 2.57  | 0.33  | 56.18 | -0.34 |
| He  | -25.25| 11.62 | 27.43 | -18.54| -23.75| 13.39 | 29.38 | -14.62|
| Li  | -15.21| -18.77| -7.88 | 8.17  | -15.09| -18.75| -7.65 | 8.15  |
| Be  | 28.46 | 31.46 | 28.58 | 37.28 | 28.65 | 31.58 | 29.20 | 38.33 |
| B   | -34.73| -36.38| -33.84| -13.20| -34.28| -35.89| -32.89| -11.51|
| C   | -18.96| -27.08| -41.74| -5.87 | -18.37| -26.26| -40.68| -4.58 |
| N   | -3.44 | -18.41| -45.27| -19.59| -3.53 | -17.99| -44.64| -21.34|
| O   | -39.94| -42.76| -37.20| -18.61| -37.04| -39.07| -33.40| -14.54|
| F   | -18.88| -22.18| -51.55| -12.10| -16.28| -18.57| -48.22| -8.40 |
| Ne  | -3.58 | -8.56 | -59.83| -10.23| -4.26 | -8.14 | -59.41| -10.42|

**Table S18.** Average basis set sensitivity, calculated as the difference in absolute errors in IPs computed with aug-cc-pV5Z and def-TZVP. Units of kJ/mol.

|         | B3LYP | PBE  | SVWN | M06  |
|---------|-------|------|------|------|
| H       | 0.75  | 0.97 | 0.84 | 0.84 |
| He      | 1.50  | 1.77 | 1.95 | 3.92 |
| Li      | 0.12  | 0.02 | 0.23 | 0.02 |
| Be      | 0.19  | 0.12 | 0.62 | 1.05 |
| B       | 0.46  | 0.50 | 0.95 | 1.69 |
| C       | 0.59  | 0.82 | 1.07 | 1.29 |
| N       | 0.09  | 0.41 | 0.63 | 1.75 |
| O       | 2.90  | 3.69 | 3.81 | 4.08 |
| F       | 2.60  | 3.61 | 3.33 | 3.70 |
| Ne      | 0.69  | 0.42 | 0.42 | 0.19 |
| Average | 0.99  | 1.23 | 1.38 | 1.85 |

Table S19. Computed electronic energies of the water molecule and water cation (in units of a.u.), and the ionization potential in units of kJ/mol, aug-cc-pV5Z basis set.

| E(B3LYP), $H_2O$ | E(B3LYP), $H_2O^+$ | IP(B3LYP) kJ/mol | E(PBE), $H_2O$ | E(PBE), $H_2O^+$ | IP(PBE) kJ/mol |
|---|---|---|---|---|---|
| -76.4369 | -75.9746 | **1213.83** | -76.3884 | -75.9235 | **1220.53** |
| E(SVWN), $H_2O$ | E(SVWN), $H_2O^+$ | IP(SVWN) kJ/mol | E(M06), $H_2O$ | E(M06), $H_2O^+$ | IP(M06) kJ/mol |
| -75.9134 | -75.4354 | **1255.16** | -76.4342 | -75.9700 | **1218.76** |

Table S20. Computed electronic energies of the water molecule and water cation (in units of a.u.), and the ionization potential in units of kJ/mol, aug-cc-pV5Z basis set, for PBE, B3LYP, and M06 using SVWN densities as trial densities.

| PBE(SVWN) | | | B3LYP(SVWN) | | | M06(SVWN) | | |
|---|---|---|---|---|---|---|---|---|
| $E_{el}$ $H_2O$ | $E_{el}$ $H_2O^+$ | IP kJ/mol | $E_{el}$ $H_2O$ | $E_{el}$ $H_2O^+$ | IP kJ/mol | $E_{el}$ $H_2O$ | $E_{el}$ $H_2O^+$ | IP kJ/mol |
| -76.3861 | -75.9209 | **1221.3** | -76.4344 | -75.97194 | **1214.08** | -76.4297 | -75.9667 | **1215.82** |

Table S21. Computed electronic energies of the water molecule and water cation (in units of a.u.), and the ionization potential in units of kJ/mol, aug-cc-pV5Z basis set, for SVWN, B3LYP, and M06 using PBE densities as trial densities.

| SVWN(PBE) | | | B3LYP(PBE) | | | M06(PBE) | | |
|---|---|---|---|---|---|---|---|---|
| $E_{el}$ $H_2O$ | $E_{el}$ $H_2O^+$ | IP kJ/mol | $E_{el}$ $H_2O$ | $E_{el}$ $H_2O^+$ | IP kJ/mol | $E_{el}$ $H_2O$ | $E_{el}$ $H_2O^+$ | IP kJ/mol |
| -75.9111 | -75.4328 | **1255.87** | -76.4368 | -75.9743 | **1214.14** | -76.4325 | -75.9685 | **1218.28** |

**Table S22.** Computed electronic energies of the water molecule and water cation (in units of a.u.), and the ionization potential in units of kJ/mol, aug-cc-pV5Z basis set, for SVWN, PBE, and M06 using B3LYP densities as trial densities.

| SVWN(B3LYP) | | | PBE(B3LYP) | | | M06(B3LYP) | | |
|---|---|---|---|---|---|---|---|---|
| $E_{el}$ $H_2O$ | $E_{el}$ $H_2O^+$ | IP kJ/mol | $E_{el}$ $H_2O$ | $E_{el}$ $H_2O^+$ | IP kJ/mol | $E_{el}$ $H_2O$ | $E_{el}$ $H_2O^+$ | IP kJ/mol |
| -75.9110 | -75.4328 | **1255.39** | -76.3884 | -75.9234 | **1220.87** | -76.4327 | -75.9688 | **1217.98** |

**Table S23.** Computed electronic energies of the water molecule and water cation (in units of a.u.), and the ionization potential in units of kJ/mol, aug-cc-pV5Z basis set, for B3LYP, PBE, and SVWN using M06 densities as trial densities.

| B3LYP(M06) | | | PBE(M06) | | | SVWN(M06) | | |
|---|---|---|---|---|---|---|---|---|
| $E_{el}$ $H_2O$ | $E_{el}$ $H_2O^+$ | IP kJ/mol | $E_{el}$ $H_2O$ | $E_{el}$ $H_2O^+$ | IP kJ/mol | $E_{el}$ $H_2O$ | $E_{el}$ $H_2O^+$ | IP kJ/mol |
| -76.4343 | -75.9726 | **1212.13** | -76.3858 | -75.9214 | **1219.28** | -75.9081 | -75.4314 | **1251.70** |

**Table S24.** Computed zero point energies from TPSSh/def2-TZVPP geometry optimization and numerically computed frequencies.

|      | ZPE (a.u.) | ZPE (kJ/mol) |
|------|-----------|--------------|
| NaCl | 0.0007976 | 2.09 |
| N2   | 0.0053607 | 14.07 |
| HF   | 0.0090684 | 23.81 |
| CO   | 0.0048478 | 12.73 |
| O2   | 0.0035435 | 9.30 |

**Table S25.** Computed bond dissociation energies (aug-cc-pV5Z basis, in kJ/mol), corrected for enthalpy and ZPE. Experimental values from the 2014 CRC Handbook of Chemistry and Physics[4].

|       | CCSD(T) | CCSD  | B3LYP  | PBE    | SVWN   | M06    | EXP.   |
|-------|---------|-------|--------|--------|--------|--------|--------|
| NaCl  | 423.8   | 414.5 | 383.2  | 392.0  | 431.9  | 427.5  | 412.1  |
| $N_2$ | 937.5   | 895.8 | 939.5  | 1005.9 | 1103.2 | 920.5  | 944.9  |
| HF    | 569.0   | 559.6 | 558.1  | 571.1  | 655.3  | 566.1  | 569.7  |
| CO    | 1075.1  | 1040.5| 1052.3 | 1113.3 | 1237.8 | 1073.9 | 1076.6 |
| $O_2$ | 493.2   | 457.3 | 506.3  | 593.4  | 722.5  | 497.0  | 498.5  |
| *MAE* | *5.3*   | *27.8*| *15.6* | *42.8* | *129.8*| *9.5*  |        |

**Table S26.** Computed electronic energies and D$_e$ (not corrected for ZPE) of diatomic molecules, aug-cc-pV5Z basis set. B3LYP, PBE, and CCSD(T).

|  | B3LYP | | PBE | | CCSD(T) | |
| --- | --- | --- | --- | --- | --- | --- |
|  | Energy (a.u.) | BDE (kJ/mol) | Energy (a.u.) | BDE(kJ/mol) | Energy (a.u.) | BDE(kJ/mol) |
| NaCl | -622.5210 | **385.3** | -622.2917 | **394.1** | -621.9901 | **425.9** |
| N$_2$ | -109.5268 | **953.6** | -109.4593 | **1019.9** | -109.4878 | **951.6** |
| HF | -100.4628 | **581.9** | -100.4022 | **595.0** | -100.4258 | **592.8** |
| CO | -113.3171 | **1065.0** | -113.2419 | **1126.0** | -113.2710 | **1087.9** |
| O$_2$ | -150.3389 | **515.6** | -150.2587 | **602.7** | -150.2657 | **502.5** |

**Table S27.** Computed electronic energies and D$_e$ (not corrected for ZPE) of diatomic molecules, aug-cc-pV5Z basis set. SVWN, M06, and CCSD.

|  | SVWN | | M06 | | CCSD | |
| --- | --- | --- | --- | --- | --- | --- |
|  | Energy (a.u.) | BDE (kJ/mol) | Energy (a.u.) | BDE(kJ/mol) | Energy (a.u.) | BDE(kJ/mol) |
| NaCl | -620.2719 | **434.0** | -622.5693 | **429.6** | -621.9756 | **416.6** |
| N$_2$ | -108.6986 | **1117.3** | -109.5137 | **934.6** | -109.4654 | **909.9** |
| HF | -99.8518 | **679.1** | -100.4647 | **589.9** | -100.4168 | **583.4** |
| CO | -112.4774 | **1250.6** | -113.3095 | **1086.6** | -113.2507 | **1053.2** |
| O$_2$ | -149.3407 | **731.8** | -150.3231 | **506.3** | -150.2434 | **466.6** |

**Table S28.** Electronic energies of diatomic molecules for PBE, B3LYP, and M06 computed using SVWN densities.

|     | PBE         |         | B3LYP       |         | M06         |         |
|     | Energy (a.u.) | BDE (kJ/mol) | Energy (a.u.) | BDE (kJ/mol) | Energy (a.u.) | BDE (kJ/mol) |
|-----|-----------|---------|-----------|---------|-----------|---------|
| NaCl | -622.2895 | 393.93  | -622.5178 | 384.67  | -622.5684 | 423.78  |
| $N_2$ | -109.4551 | 1019.71 | -109.5224 | 952.40  | -109.5077 | 937.13  |
| HF  | -100.3998 | 596.26  | -100.4603 | 583.57  | -100.4572 | 583.90  |
| CO  | -113.2379 | 1126.90 | -113.3124 | 1064.49 | -113.3026 | 1086.40 |
| $O_2$ | -150.2542 | 603.22  | -150.3334 | 514.07  | -150.3150 | 507.08  |

**Table S29.** Electronic energies of diatomic molecules for SVWN, B3LYP, and M06 computed using PBE densities.

|     | SVWN        |         | B3LYP       |         | M06         |         |
|     | Energy (a.u.) | BDE (kJ/mol) | Energy (a.u.) | BDE (kJ/mol) | Energy (a.u.) | BDE (kJ/mol) |
|-----|-----------|---------|-----------|---------|-----------|---------|
| NaCl | -620.2691 | 433.84  | -622.5204 | 384.68  | -622.5733 | 426.80  |
| $N_2$ | -108.6946 | 1117.53 | -109.5261 | 953.15  | -109.5114 | 936.49  |
| HF  | -99.8495  | 680.83  | -100.4626 | 582.07  | -100.4601 | 583.89  |
| CO  | -112.4735 | 1251.77 | -113.3162 | 1064.20 | -113.3068 | 1088.80 |
| $O_2$ | -149.3362 | 732.59  | -150.3378 | 514.02  | -150.3196 | 507.31  |

**Table S30. Electronic energies of diatomic molecules for SVWN, PBE, and M06 computed using B3LYP densities.**

|     | SVWN | | PBE | | M06 | |
| --- | --- | --- | --- | --- | --- | --- |
|     | Energy (a.u.) | BDE (kJ/mol) | Energy (a.u.) | BDE (kJ/mol) | Energy (a.u.) | BDE (kJ/mol) |
| NaCl | -620.2685 | **433.49** | -622.2914 | **393.61** | -622.5748 | **428.61** |
| $N_2$ | -108.6942 | **1116.27** | -109.4585 | **1019.22** | -109.5110 | **932.53** |
| HF | -99.8494 | **681.07** | -100.4019 | **595.00** | -100.4607 | **583.46** |
| CO | -112.4727 | **1250.34** | -113.2410 | **1125.26** | -113.3073 | **1087.14** |
| $O_2$ | -149.3354 | **730.82** | -150.2578 | **601.46** | -150.3195 | **503.52** |

**Table S31. Electronic energies of diatomic molecules for SVWN, B3LYP, and PBE computed using M06 densities.**

|     | SVWN | | B3LYP | | PBE | |
| --- | --- | --- | --- | --- | --- | --- |
|     | Energy (a.u.) | BDE (kJ/mol) | Energy (a.u.) | BDE (kJ/mol) | Energy (a.u.) | BDE (kJ/mol) |
| NaCl | -620.2624 | **430.49** | -622.5178 | **385.65** | -622.2875 | **392.85** |
| $N_2$ | -108.6922 | **1119.87** | -109.5237 | **950.28** | -109.4565 | **1019.77** |
| HF | -99.8463 | **678.79** | -100.4607 | **580.43** | -100.3995 | **593.85** |
| CO | -112.4694 | **1248.81** | -113.3137 | **1063.01** | -113.2380 | **1125.30** |
| $O_2$ | -149.3333 | **734.22** | -150.3357 | **514.36** | -150.2558 | **605.50** |

Table S32. Computed Co-C bond dissociation energies (kJ/mol) of 5'-deoxyadenosylcobalamin (def2-TZVPP basis, not corrected for zero-point energies).

|  | E(B3LYP) | E(PBE) | E(SVWN) | E(M06) |
|---|---|---|---|---|
| $\rho_{B3LYP}$ | 65.9 | 123.1 | 199.2 | 121.1 |
| $\rho_{PBE}$ | 69.5 | 119.8 | 200.8 | 117.0 |
| $\rho_{SVWN}$ | 67.3 | 120.6 | 200.0 | 119.0 |
| $\rho_{M06}$ | 68.2 | 117.1 | 198.1 | 102.4 |
| SENSITIVITY (MAD from native) | 2.4 | 3.9 | 1.2 | 8.3 |